\documentstyle[12pt,aasms4]{article}
\slugcomment{}
\lefthead{Enya et al.}
\righthead{$JHK'$ variability of AGNs}

\begin{document}

\title{$JHK'$ Imaging Photometry of Seyfert 1 AGNs and Quasars II: \\
            Observation of Long-Term Variability}

\author{Keigo Enya\altaffilmark{1}, Yuzuru Yoshii\altaffilmark{1,4},
   Yukiyasu Kobayashi\altaffilmark{2}, Takeo Minezaki\altaffilmark{1}, 
   Masahiro Suganuma\altaffilmark{3}, Hiroyuki Tomita\altaffilmark{3}
   and Bruce A. Peterson\altaffilmark{5}}

\vspace{20mm}

\affil{$^1$ Institute of Astronomy, School of Science,
              University of Tokyo, Osawa 2-21-1,  
              Mitaka, Tokyo 181-8588, Japan}
\affil{$^2$ National Astronomical Observatory,
              Osawa 2-21-1, Mitaka, Tokyo 181-8588, Japan}
\affil{$^3$ Department of Astronomy, University of Tokyo,
             Hongo 7-3-1, Bunkyo-ku, Tokyo  113-0033, Japan}
\affil{$^4$ Research Center for the Early Universe (RESCEU),
          School of Science, University of Tokyo,
             Hongo 7-3-1, Bunkyo-ku, Tokyo 113-0033, Japan}
\affil{$^5$ Research School of Astronomy and Astrophysics, The
         Australian National University, Weston Creek, ACT 2611,
         Australia}


\begin{abstract}

Observations of 226 AGNs in the near-infrared $J$, 
$H$, and $K'$ bands are presented along with the analysis of the
observations for variability.  Our sample consists 
mainly of Seyfert 1 AGNs and QSOs. About a quarter of the objects in 
each category are radio loud.   
The AGNs in the entire sample have the redshifts spanning the range from 
$z=0$ to 1, and the absolute magnitudes from $M_B=-29$ to $-18$.  All 
the objects were observed twice and their variability was measured by 
differential photometry.

A reduction method of differential photometry, optimized to the analysis 
of extended images, has been developed.   
The systematic error in variability arising from AGNs of highly extended 
images is estimated to be less than 0.01 mag in each of the $J$, $H$, 
and $K'$ bands.  The systematic error arising from the flat 
fielding is negligible for most AGNs, although it is more than 0.1 mag for 
some particular cases. The overall average flat fielding error is 
0.03 mag for the image pairs.  
We find that these systematic errors are superseded by statistical 
errors, and the overall average total 
 systematic and statistical errors amounts to 0.05
mag in the measured variability in each band.

We find that 58\% of all the AGNs in the entire sample show variability 
of  more than $2\sigma$, and 44\% of more than $3\sigma$. This 
result holds independent of the $J$, $H$, and $K'$ bands.  The detection 
rate of variability is higher for a subsample of higher photometric 
accuracy, and there appears no limit to this tendency.  
In particular, when we consider a subsample  with small photometric 
errors of $\sigma<0.03$ mag, the rate of $2\sigma$ detection is 80\%, 
and 64\% for $3\sigma$ detection.  This suggests that most AGNs are 
variable in the near-infrared.

\end{abstract}



\section{Introduction}

Active Galactic Nuclei (AGNs) emit enormous amount of 
energy from their central, 
compact region.  However, it is impossible to resolve this compact region 
with current instruments when the AGNs are at cosmological distances.  
Observation of variability for such distant objects is a useful 
method for investigating the  emitting mechanism as well as the  internal 
structure of the compact region.  Therefore, a first pathway to progress 
is to detect their variability in various passbands and to study 
the detected variability in detail.  In this regard, it is essential 
to measure not only the magnitude of variability but also its error 
as accurately as possible.

Recent monitoring observation of AGNs by differential photometory with 
CCDs detected variability in almost all of the AGNs observed in 
the optical bands (Borgeest \& Schramm 1994, Netzer et al. 1996, Giveon 
et al. 1999).  Determination of near-infrared (NIR) variability in
AGNs, for 
which optical variability has already been detected, leads to a better 
understanding of their emission mechanisms. In the dust 
reverberation model, the  structure of the dust torus can be investigated 
from 
the time delay of the NIR variability, which lagges behind the 
UV/optical variability (Cravel, Wamsteker \& Glass 1989, 
Barvanis 1992).  

Neugebauer et al. (1989) monitored 108 optically selected quasars in the
$J$, $H$, $K$, $L$ and 10$\mu$m bands, and discussed the  probability of 
each AGN having varied.  However, the number of AGNs 
with NIR variability measurements of
comparable accuracy to the optical 
study, is still too small to derive general properties of variability. 
Therefore, it is of prime importance at this stage to obtain more data
of high accuracy on NIR variability.  

Here, in this paper, we present the new $JHK'$ variability data for  226 
AGNs. To date, this is the largest sample
systematically ovserved in the NIR.  Each AGN was observed twice. 
Efficient and highly accurate detection of variability 
was achieved using differential photometry.  
The method  of error estimation and the generality of 
the phenomenon of variability are discussed 
in this paper.  Various relations among detected variability, absolute 
$B$-magnitude and AGN redshift will be discussed in Paper III.

We have been conducting a project called MAGNUM which
is an acronym for  Multicolor Active Galactic NUclei Monitoring
(Kobayashi et al. 1998a, 1998b).  
The MAGNUM Project aims to determine the distances to many AGNs from  
measurements of the time delay between NIR and UV/optical valiabilities.  
This paper is therefore a preliminary  target selection of   
AGNs with sizable variability, which will be monitored by the MAGNUM 
Project.

\section{Observations}

The data analyzed in this paper was obtained using the image reduction 
processes 
described in \S 2 of Paper I.  We selected
226 AGNs from the calalogs of Quasars and Active Galactic Nuclei 
(Veron-Cetty and Veron 1993, 1996, 1998), which will also be used as target 
AGNs for the MAGNUM Project.  Their right ascensions 
and declinations are shown in Fig.1 of Paper I, and 
their redshifts and absolute $B$-magnitudes in Fig.2 of Paper I.

All observations were performed with the NIR camera 
PICNIC (Kobayashi et al.1994) mounted on the $1.3$m infrared 
telescope at the Institute of Space and Astronomical Science (ISAS), 
Japan.  AGNs were imaged by stepping the telescope in a raster pattern. 
Two standard stars at different altitutes were observed three 
times each night in the $J$, $H$ and $K'$ bands.  
In usual cases, images were reduced automatically by the software PICRED 
developed for the PICNIC camera and optimized for the  reduction of 
AGN and QSO images. 

\section{Analysis}

\subsection{Differential photometry for detecting the variability of AGNs}

\subsubsection{The problems of standard star-based photometry}

Each of AGNs was observed on two different nights 
with photometric standards.  However, some problems arise
when one wishes to regard the difference between two measured
magnitudes as variability.

One of the problems arises from different the PSFs for AGNs and 
the photometric standards which occur due to variation of seeing 
and/or telescope tracking.  Adopting a larger aperture alleviates
this problem, but at the same time it decreases the  S/N ratio of 
an object image.  
The S/N ratio increases by increasing the integration 
time, but longer integration time obviously 
decreases the number of observable objects per night.

Another problem arises from the different atmospheric transmissivities 
at the different times when the AGNs and the photometric standards
were observed.  
Frequent observation of photometric standards alleviates this problem, 
but at the same time it decreases the observational efficiency.  
The atmospheric transmissivity as well as the seeing conditions could 
vary significantly, especially when a cloud comes into the line of 
sight. Great care was taken to avoid such interviening clouds during 
the observations, but some very thin clouds were identified only after 
daybreak. The data taken on such nights and those on non-photometric 
nights were excluded from the analysis, which also leads to a loss 
of observational efficiency.  

The above problems stem from magnitudes of images taken 
on different nights, along different lines of sight, and under different 
instrumental conditions such as the  tracking of the telescope.  Unless these 
problems are avoided, the standards-based photometry is 
not an accurate  method for detecting the variability of AGNs.  

\subsubsection{The superiority of differential photometry for variabilty 
               detection}

In principle, the above problems from variations of PSF and 
transmissivity are minimized by differential photometry.  
Therefore, we adopted a method of differential 
photometry in order to detect the variability of 
AGNs with high accuracy and reliability.  
In the ideal  case such that target AGNs and nearby 
reference objects are point sources and no errors exist in flat 
fielding, we describe the scheme of differential photometry 
which is used to detect the variability of AGNs.

We measure instrumental aperture magnitudes of both AGNs and 
reference objects on two different nights.  We regard their 
magnitude difference as reflecting real difference of their 
fluxes, because the systematic errors
from variations of seeing and telescope tracking
 are cancelled out under the assumptions that the 
PSFs in each field are the same within the required accuracy and that 
variation of atmospheric transmissivity  equally affects 
all objects in each field.  
We therefore regard the change of measured magnitudes from  
different nights for each AGN as its variability.  Cancellation of 
systematic errors in the case of differential photometry justifies 
our use of a smaller aperture, which results in smaller statistical 
errors as well.  It is worth mentioning that a high observational 
efficiency is maintained by differential photometry, because 
the data taken through thin clouds can also be used.

There remain some systematic errors that are not suppressed even 
by differential photometry.  Imperfect of flat fielding 
gives rise to errors in the  magnitude difference between an AGN 
and its nearby reference objects.  Variation in the  PSF is not completely 
canceled if intrinsic brightness profiles are not the same.  
Furthermore, the possible variability of the reference objects, of course, 
is  another source of error in determination of the valiability 
of the AGN.

However, as describe below, all these errors were not fatal in our 
analysis, where the images of AGNs are not very extended and most 
reference objects consist of stars.

\subsubsection{The use of DSS images as positional reference}

Identification of AGNs and reference objects observed on two 
different nights is an important step in  differential photometry.  
We used images from a digitized Palomer chart, or 
equivalently, the Digitized Sky Survey (DSS), as positional references. 
The identification of 
each object on three frames, such as a DSS frame and two NIR 
frames taken on  different nights, is more reliable than only on 
two NIR frames.

The DSS images of all AGNs observed in this paper were 
obtained with a batch script from a database at the Data Analysis 
Center of the National Astronomical Obvervatory, Japan (NAOJ); 
DDS2 images are of better resolution compared to those of 
DSS1.  We use as many DSS2 images as possible, while not 
excluding the use of DSS1 images for the purpose of completeness.  
The size of the extracted DSS images was $12\times 12$ arcmin$^2$.
That size is larger 
than the mosaiced NIR images we obtained.

Objects in the DSS images were automatically detected by the 
IRAF cl script which consists of the DAO find task and a perl script.
After some trial-and-error tuning of the detection parameters, we
found that objects are distinguished from noise more easily 
in the DSS images than in our  NIR images.

Associating each target AGN with a DSS image was done manually.
The target was mostly located at the center, with some 
exceptions where it was  a little away from the center owing to positional 
errors. The finding charts were prepared based on the information 
in the VV catalog.  For efficient association, the cl script was 
prepared and used in such a way that the nearest object from 
mouse-clicked point was searched.  The found object was, if 
close enough, regarded as the target AGN, otherwise an error 
message of `not found' was given. The result of 
detection/non-detection of AGNs and nearby reference objects 
were written in a text file within  other additional information.

\subsubsection{The detection of objects in NIR images}

Objects in the NIR images were automatically detected using 
the cl script similarly to the case for the DSS images.  
Luminosity of our AGNs is distributed over a wide range, so 
that a large number of noise images  were detected when the threshold 
was  set low enough to detect faint AGNs.  In such a case,
many false detections emerge near the faint end of the detected 
object distribution.  An appropriate level of faint object detections
was estimated and the 
faint object distribution  was automatically smoothed with the IRAF task 
for image 
replacement.  By this operation, false detections were greatly 
decreased, though not completely excluded.

Together with nearby reference objects, most AGNs were detected 
in the automatic detection scheme, while some of faint AGNs required 
manual trial-and-error tuning of the detection parameters.  However, 
invisible AGNs, or very faint AGNs that were not detected by any 
manual method were regarded as non-detections.

The most central object was not always the target AGN in the NIR 
images, partly because the tuning of the telescope pointing was 
not done for some time.  Furthermore, since the fourth quadarant of
the detector was out of order in the second of the two 
observation perios, an off-center region of the detector was used to
detect the AGNs in the second periods.  
Therefore it was necessary to identify and mark the 
target AGNs manually using the finding charts.

\subsubsection{The identification of reference objects}

It was necessary to identify reference objects automatically, 
because their number exceeds that of AGNs by about an order 
of magnitude
and is beyond ability of manual operation.  The DSS and NIR 
images were listed downwards from the top, and  arranged according to 
their coordinates from north to south.  The coordinates of 
detected objects or AGNs in such images were then transformed 
to the new coordinates in units of arcsec, placing each AGN at 
the origin. Identification between the DSS and infrared images 
was performed by a usual least-squares fitting with two free 
parameters of rotation and expansion of the images.   
At the first step of the fitting,
reference objects in the NIR images and those in the DSS 
images, if within a seperation of 10 arcsec, were regarded as 
the same objects.

The fitting was iterated 
with a smaller aperture using  new starting parameters which were 
obtained as the result of the last iteration. The lower limit 
for  decreasing the aperture was set to be 2 arcsec, which is 
comparable to a typical FWHM for a  point source in the NIR 
images.  The final value of rotation parameter was less than a 
few degrees, and that of the expansion factor was about $1 \pm 0.05$ 
for most cases.  The automatic fitting was almost always 
successful, and the script produced from the last iteration was 
useful.  For some cases where only very faint objects were 
detected in the images, the iteration did not show a tendency 
of convergence,  so that  manual operation was needed to tune 
the parameters.  However, if even manual identification, 
the  data were not used in our analysis.  Only the pairs of 
DSS and NIR objects, identified through the above fitting 
procedure, were regarded as the same objects.

\subsubsection{The estimation of AGN variability}

In order to derive the variability of AGNs from the difference 
between their magnitudes observed on  different nights,  
it is necessary to determine the fiducial which should be 
subtracted from such difference. 
Let $\Delta m_i =m_{i, 2}-m_{i, 1}$ be the difference between 
the instrumental magnitudes of the $i$-th reference object 
observed on two  different nights, and $\sigma_{\Delta m_i}^2
=\sigma_{m_{i, 2}}^2+\sigma_{m_{i, 1}}^2$ its measurement 
errors, where $i$ runs from 1 to $N$.  The subscript 
1 or 2 refers to  the first or second measurement of the  two observations, 
respectively.

The distribution of $\Delta m_i$ is broader than expected 
from $\sigma_{\Delta m_i}$. This broadening likely stems from 
an  error in the  flat fielding which would equally affect  all 
reference objects regardless of their luminosity. 
We incorporate this extra error $\sigma_f$ in the estimation of 
the total error, so that the distribution of $\Delta m_i$ 
conforms to $\chi^2$ distribution.  
The value of $\sigma_f$ is then obtained from 
solving the following equation:  
\begin{equation}
\sum_{i} \frac{(\Delta m_i -\Delta m)^2}
    {\sigma_{\Delta m_i}^2+\sigma_f^2}=N-1  \;\;, 
\end{equation}
where $\Delta m$ is a weighted average given by
\begin{equation}
\Delta m={\sum_i \frac{\Delta m_i}
                 {\sigma_{\Delta m_i}^2+\sigma_f^2} }
    /{\sum_i \frac{1}{\sigma_{\Delta m_i}^2+\sigma_f^2 }} \;\;,
\end{equation}
where we set $\sigma_f=0$ if no real value exists. Substitution 
of the value of $\sigma_f$ into equation 2 gives the value of 
$\Delta m$ which should be used as a fiducial in 
deriving the variability of AGNs.  The uncertainty of this 
fiducial is estimated as
\begin{equation}
\sigma_{\Delta m}^2=
  \frac{1}{N-1} \sum_i 
  \frac{(\Delta m_i-\Delta m)^2}
  {\sigma_{\Delta m_i}^2+\sigma_f^2}
  /{\sum_i \frac{1}{\sigma_{\Delta m_i}^2+\sigma_f^2}} \;\;.
\end{equation}
The variability of AGN is therefore given by
\begin{equation}
\Delta m_{\rm AGN}=m_{{\rm AGN}, 2}-m_{{\rm AGN}, 1}- \Delta m \;\; ,
\end{equation}
and its error by
\begin{equation}
\sigma_{\Delta m_{\rm AGN}}^2=\sigma_{m_{{\rm AGN}, 2}}^2 
     + \sigma_{m_{{\rm AGN}, 1}}^2 
     + \sigma_{\Delta m}^2   \;\; .
\end{equation}

The process of determining the fiducial $\Delta m$ 
is iterated, by excluding abnormal reference objects beyond 
a deviation  of 5$\sigma$ deviation.  Three iterations suffice to exclude all 
such abnormal objects which are mostly either 
misidentifications, or near the edge of frame, 
or intrinsicly variable stars. After convergence at $5\sigma$,
the iterative rejection is repeated at $3\sigma$.

An example of differential photometry is shown in Fig. 1, 
where the difference $m_{{\rm AGN}, 2}-m_{{\rm AGN}, 1}$ for 
Mark 1320 in the $H$ band is plotted at $i=0$, together with 
$\Delta m_i$ for five reference objects from $i=1$ to 5.  
It is evident that the fiducial $\Delta m$ (dashed horizontal 
line) yields $\Delta m_{\rm AGN}=-0.55$ mag, that is, Mark 1320 
was brightened by 0.55 mag.

\subsection{The effect of PSF variability}

The determined variability, $\Delta m_{\rm AGN}$, of AGNs by 
differential photometry is less sensitive to PSF variation 
as  compared to that by standards-based photometry. 
In principle, the effect of PSF variation 
on differential photometry is canceled out, 
if the PSFs of target and reference 
objects are of the same shape. However, since the target 
AGNs have more extended profiles than reference objects 
that are mostly stars, the effect of PSF variation is not 
completely canceled out.  
The variability of AGNs, corrected for the effect of PSF 
variation, is therefore given by
\begin{equation}
\Delta m_{\rm AGN}=m_{{\rm AGN}, 2}-m_{{\rm AGN}, 1}
 - \Delta m - \Delta m_{\rm PSF} \;\; ,
\end{equation}
and its error by
\begin{equation}
\sigma_{\Delta m_{\rm AGN}}^2=\sigma_{m_{{\rm AGN}, 2}}^2 
     + \sigma_{m_{{\rm AGN}, 1}}^2  
     + \sigma_{\Delta m}^2   
     + \sigma_{\Delta {\rm PSF}}^2  \;\; .
\end{equation}

We characterize the PSFs of the reference objects by their 
FWHMs measured with the cl script in the IRAF imexamine task.  
There is a tendency for the  FWHMs from the IRAF output to be  
larger for fainter objects.  In order to avoid this tendency, 
we use only the reference objects for which photon counts at 
their profile center exceed 50.  We estimate the difference 
$\Delta {\rm FWHM}_i ={\rm FWHM}_{i, 2}-{\rm FWHM}_{i, 1}$ for 
the $i$-th reference object around the target AGN, where the 
subscript 1 or 2 refers to the first or second measurement 
from the two 
observations, respectively.  We then derive the correction 
$\Delta m_{{\rm PSF}, i}$ due to an intrinsic 
nonzero  $\Delta {\rm FWHM}_i$, 
by convolving the smaller-FWHM$_i$ image to match the 
larger-FWHM$_i$ image.  This analysis was performed using the 
IRAF gaus task, and the correction $\Delta m_{{\rm PSF}, i}$ 
is derived as a function of $\Delta {\rm FWHM}_i$.  Taking an 
average over all the reference objects around the target AGN, 
we obtain $\Delta m_{\rm PSF}$ and $\Delta {\rm FWHM}$, which 
will be used to correct $\Delta m_{\rm AGN}$ for the effect of 
a nonzero intrinsic PSF.  

The top panel of Fig. 2 shows the correction $\Delta m_{\rm PSF}$ 
plotted against $\Delta {\rm FWHM}$, based on the data for which 
the accuracy of instrumental magnitude is better than 0.02 mag.
It is clear that the better (poorer) seeing in the later of 
two observations gives fake darkening (brightening) in 
$\Delta m_{\rm AGN}$. The plots in this panel are fitted by a 
linear relation between $\Delta m_{\rm PSF}$ and 
$\Delta {\rm FWHM}$, and the dispersion around 
$\Delta m_{\rm PSF}$ is expressed, in terms of 
$\Delta {\rm FWHM}$, as
\begin{equation}
  \sigma_{\Delta m_{\rm PSF}}^2
  =\left( \frac{d\Delta m_{\rm PSF}}
 {d\Delta {\rm FWHM}} \right)^2 
  \sigma_{\Delta {\rm FWHM}}^2 \;\; .
\end{equation}
The bottom panel shows the frequency distribution of $\Delta m_{\rm PSF}$.
Since the data are restricted to brighter AGNs of high accuracy, we 
cannot individually correct the variability of AGNs in our sample.  
With this restriction, we rather estimate $\Delta m_{\rm PSF}\approx 0$ 
and $\sigma_{\Delta m_{\rm PSF}}\approx 0.01$ from the combined $JHK'$ 
data and substitute these values in equations 6 and 7, irrespective of 
the passband.

Here, the estimate of 
$\Delta m_{\rm PSF}$ has been limited to bright AGNs only. 
However, contrary to the case of bright AGNs, the fake variability 
due to PSF variation would be even smaller for fainter AGNs, 
because only the central, point-like part of their profiles is 
visible with their extended hosts mostly below detection.  Therefore, 
the estimate of $\Delta m_{\rm PSF}\approx 0$ from bright AGNs
is also used to correct the variability of fainter AGNs in this paper.

It is worth mentioning that this paper is a preliminary study for a monitoring
program, MAGNUM,  which is expected to go fainter 
($R_{lim}\sim 23$mag) and more accurately 
($\sigma\sim 0.01$mag), compared with our observations
here.  In such a program, the correction factor for each of 
AGNs is obtained as a function of varying luminosity and 
PSF in the course of lon-term monitoring observations, and the  method 
of convolving obtained images, as used here, may not be  necessary.

\subsection{The test of the error estimation}\label{ssec_seido_hyouka}

Error estimation in the variability data is very important.  
When the error is sufficiently small, a fake variability
originates  from only the dispersion of the photometric data. 
On the other hand, when the error is too large, 
the significance of any detected  
variability cannot be certain. 

The variability of reference objects was fairly small, because 
most of them are stars.  Their corrected variability is denoted 
by $\Delta m_{c, i}=\Delta m_i-\Delta m$, and the corrected error 
is calculated by $\sigma_{\Delta m_{c, i}}({\rm cal})^2=
\sigma_{\Delta m_i}^2+\sigma_f^2$.  We divide the reference 
objects 
into subsamples according to their calculated errors binned at 
intervals of 0.01.  For the subsample in each error bin, we have 
constructed the frequency distribution of corrected variability, and 
derived the median $\Delta m_c$ and dispersion 
$\sigma_{\Delta m_c}({\rm dsp})$ for a sample in each error bin 
centered at $\sigma_{\Delta m_c}({\rm cal})$.  

Figure 3 shows $\Delta m_c$ and $\sigma_{\Delta m_c}({\rm dsp})$ 
plotted against $\sigma_{\Delta m_c}({\rm cal})$ in the $J$, $H$, 
and $K'$ bands.
It is seen from this figure that the $\sigma_{\Delta m_c}({\rm cal})$'s
are nearly equal to the $\sigma_{\Delta m_c}({\rm dsp})$'s, but are 
mostly below the dotted line of $\sigma_{\Delta m_c}({\rm cal})=
\sigma_{\Delta m_c}({\rm dsp})$.  This indicates a slight 
overestimation 
of $\sigma_{\Delta m_c}({\rm cal})$, which is desirable rather than 
the converse, so that our error estimation of differential 
photometry is quantitatively reliable over a range of error from 0.01 
up to 0.1 for the $J$, $H$ and $K'$ bands.
The medians, $\Delta m_c$'s, are almost equal to zero.

\subsection{A comparison of  the accuracies of two photometry methods}

In this paper we have used a method of differential photometry to detect 
the NIR variability of AGNs. On the other hand, however, we 
can also use a method of standards-based photometry to detect AGN 
variability in our sample.  In this section we compare the accuracies 
of these two methods when applied to the same AGNs.

The differential photometry uses the instrumental, aperture magnitudes 
with a fixed aperture of $r=3$ pixels, while the standards-based 
photometry uses  aperture magnitudes with  $r=7, 10, 12$ and 15 
pixels.  By standards-based photometry we obtained a pair of magnitude and 
measurement error for each object observed on two different nights, 
and determined the magnitude difference $\Delta m_i=m_{i, 2}-m_{i, 1}$ 
and its error $\sigma_{\Delta m_i}^2=\sigma_{m_{i, 2}}^2 +
\sigma_{m_{i, 2}}^2$.  The analysis below is limited to the data for 
which more than two reference objects were used for differential 
photometry.  

Figure 4 shows the comparison between the variabilities of AGNs by 
differential photometry $\Delta m_{\rm AGN} ({\rm dif})$ and by 
standards-based photometry $\Delta m_{\rm AGN} ({\rm std})$ for 
the case of the $J$ band.  Similar results in the $H$ and $K'$ bands 
are not shown here.  The distribution of data points, for the 
standards-based photometry with $r=7$ pixels, is elongated, more or 
less, along the diagonal in the $\Delta m_{\rm AGN} ({\rm dif})$ 
versus $\Delta m_{\rm AGN} ({\rm std})$ diagram. However, 
increasing the aperture from $r=7$ to 15 pixels steadily increases the 
error (see below), which destroys a clear correlation between 
$\Delta m_{\rm AGN} ({\rm dif})$ and $\Delta m_{\rm AGN} ({\rm std})$.

Figure 5 shows the comparison between errors by 
differential photometry $\sigma_{\Delta m_{\rm AGN}} ({\rm dif})$ and by 
standards-based photometry $\sigma_{\Delta m_{\rm AGN}} ({\rm std})$ 
for the case of $J$ band.  Similar results in the $H$ and $K'$ bands are
not shown here.  The first feature to 
be noticed is that many data points are concentrated along a 
linear sequence in the $\sigma_{\Delta m_{\rm AGN}} ({\rm dif})$ 
versus $\sigma_{\Delta m_{\rm AGN}} ({\rm std})$ diagram. 
The second feature is that the sequence is inclined much closer 
to the $\sigma_{\Delta m_{\rm AGN}} ({\rm std})$-axis for larger 
aperture. These features 
are consistent with a tendency in aperture photometry such that 
the S/N ratio becomes smaller for larger aperture,  enhancing 
$\sigma_{\Delta m_{\rm AGN}} ({\rm std})$.

The linear sequence seen in the $\sigma_{\Delta m_{\rm AGN}} ({\rm dif})$ 
versus $\sigma_{\Delta m_{\rm AGN}} ({\rm std})$ diagram always has a 
slope much less than 1.  In particular, the slope is about 0.5 for the 
aperture of $r=7$ pixels, and about 0.1 for $r=15$ pixels.  
This shows that the accuracy of differential photometry is superior 
to standards-based photometry. 

A small number of outliers exist far above the linear sequence.  
Their existence may be explained partly by larger fiducial error 
in differential photometry which gives a larger 
$\sigma_{\Delta m_{\rm AGN}} ({\rm dif})$ even 
for higher S/N ratio.  Such fiducial error may stem from a possible 
error of flat fielding, a contamination by variable stars in reference 
objects, and inaccurate photometry of objects near the edge of the
frame.  Because of their small number, exclusion of such outliers
from the sample does not improve the statistics in this paper.

Next, we quantitatively compare the accuracies of two methods, 
based on a frequency distribution of AGNs with respect 
to an error ratio $c = \sigma_{\Delta m_{\rm AGN}} ({\rm std})/
\sigma_{\Delta m_{\rm AGN}} ({\rm dif})$.  Here the 
data plotted in Fig. 5 are used.  Let $f(\ge c)$ be a fractional 
number of AGNs, having an error ratio larger than $c$, 
relative to their total number.  Figure 6 shows the fraction 
$f(\ge c)$ plotted as a function of aperture for various 
values of $c$ in the $J$, $H$, and $K'$ bands.

This figure indicates that for most cases the fractional number of AGNs 
with the error ratio beyond unity comprises about $f=80\%$ in the sample.  
For standards-based photometry, the aperture which is 3 or 4 
times larger than typical FWHM is usually used to avoid the 
influence of PSF variation. This corresponds to an aperture of 
$r=10$ or 12 pixels, because a typical FWHM is about 3 pixels in our 
observations.  Such apertures give $c\approx 2-3$ for $f=50\%$, 
as seen from Fig. 6.  Therefore, the accuracy of 
differential photometory is $2-3$ times higher than that of  
standards-based photometry.

Further suppression of fiducial error in differential photometry 
is possible, if the effects of flat-fielding error and inaccurate 
photometry near the frame edge can be minimized.  It is then hoped 
that many bright reference objects can be chosen close enough to the 
target AGN, but not too close to produce  confusion.  In such an ideal
situation, the photometry only in some relevant area of all the dithering 
frames suffices, in contrast to the photometry over the entire  
dithered area as used in this paper.  

In order for the MAGNUM Project to realize much higher accuracy in 
differential photometry, we select the candidate AGNs, which should 
satisfy the above criterion. Owing 
not only to the sample selection, but also considering exposure time 
and dithering pattern, we can make the analysis go fainter without 
degrading the S/N ratio.

\section{Results}

A certain fraction of AGNs have certainly varied in the $J$, $H$ 
and $K'$ bands.  In this paper,  
we consider a sample of 226 AGNs data obtained with 
more than two reference objects and with an  accuracy better than 
0.1 mag.  Their variability data  
are tabulated in Table 1.  We show the distribution of variability 
$\Delta m_{\rm AGN}$ and redshift $z$ for the entire sample in 
Fig. 7, and the histogram with respect to $\Delta m_{\rm AGN}$ 
in Fig. 8.

As a measure of this variability, we introduce a ratio 
$R=\Delta m_{\rm AGN}/\sigma_{\Delta m_{\rm AGN}}$, and 
construct the frequency distribution of AGNs with respect to $R$.  
Figure 9 shows such a frequency distribution for our
entire sample of 226 AGNs, and two samples of radio-quiet and
radio-loud AGNs. All AGNs with $R>5$ are those that have varied 
most certainly and are included in the rightmost bin of $R=5-6$.  
For both the entire sample and radio-quiet sample, the frequency 
monotonically decreases rightwards from the bin of $R=0-1$ to 
$R=4-5$, and  the peak occurs in the rightmost bin of $R=5-6$.  
However, because of the small sample size, it is not easy to see 
similar features for the radio-loud sample. These results hold 
commonly for the $J$, $H$ and $K'$ bands.  

Figure 10 shows the frequency distribution for the 
radio-quiet and radio-loud AGNs which are furthermore classified 
by rest-frame time interval $\Delta t_{\rm rest}$ between our two 
observations made for each AGN.  For the 
radio-quiet AGNs with short time interval of 
$\Delta t_{\rm rest}=100-400$ days, the frequency monotonically 
decreases with increasing $R$, and the frequency for $R>5$ is rare.  
On the other hand, for the radio-quiet AGNs, with the long time 
interval of $\Delta t_{\rm rest}=400-800$ days, the frequency stays 
almost constant over a range of $R=0-5$, and shows a prominent peak 
at $R>5$. Therefore, most of radio-quiet AGNs that have certainly 
varied are those with time interval exceeding 400 days in rest frame.  
For the radio-loud AGNs, the frequency stays, more or less, constant, 
independent of time interval, though the sample size is too small to 
claim it definitely.

Figure 11 shows the frequency distribution for all 
AGNs which are divided by their error into three groups of high 
accuracy of $\sigma_{\Delta m_{\rm AGN}}<0.03$, intermediate 
accuracy of $\sigma_{\Delta m_{\rm AGN}}=0.03-0.05$, and  
low accuracy of $\sigma_{\Delta m_{\rm AGN}}=0.05-0.1$.  
It is seen from this figure that the frequency distributions for these 
three groups differ remarkably from each other.  For the high-accuracy
group, the frequency is localized in the rightmost bin of $R>5$, 
indicating that most AGNs in this group have certainly varied.  On the 
other hand, for the low-accuracy group, the frequency is  enhanced on 
the left.  The frequency distribution for the intermediate-accuracy 
group is in between these two extremes.  

Table 2 tabulates the estimated fraction of AGNs 
that have certainly varied with $2\sigma$ or $3\sigma$ confidence in 
various cases of different samples.  The fraction of varied AGNs 
with $2\sigma$ confidence in the high-accuracy group is 85\%($J$), 
82\%($H$), and 73\%($K'$), and the fraction for $3\sigma$ confidence 
is 73\%($J$), 67\%($H$), and 65\%($K'$). Therefore, based on the 
high-accuracy group with $\sigma_{\Delta m_{\rm AGN}}<0.03$, we at 
least conclude that the detection rate of variable AGNs is about 
80\%($2\sigma$) or 68\%($3\sigma$).   

If the sample indeed contains some non-variable AGNs, the fraction 
of variable AGNs should become saturated as the accuracy increases.
Comparing three accuracy groups, we see no sign of such saturation 
towards higher accuracy.  However, it is not possible to extrapolate 
this tendency ariving eventually at 100\% to much higher accuracy, 
because other parameters are not uniform in the three accuracy groups. 
In particular, the high-accuracy group is biased in favor of nearby,
bright, and radio-quiet AGNs.

In contrast with our higher detection rate, Neugebauer et al. (1989) 
reported that about 24\% of 108 AGNs in their $K$-band monitoring 
sample have certainly varied at the 99.7\% confidence level, which 
is comparable to our result based on the low-accuracy 
group with $\sigma_{\Delta m_{\rm AGN}}=0.05-0.1$.  Moreover, the shape of 
frequency distribution with respect to $R$ by Neugebauer et al. 
(1989) is also consistent with our result based on the 
low-accuracy group.  
Consequently, it is likely that 
their low detection rate is only apparent because of the rather 
low accuracy in their observations which were based on the single detector 
with a $5-15$ arcsec beam. 

\section{Conclusion}

We present the $JHK'$ variability data for a sample of 
226 AGNs consisting mainly of Seyfert 1 AGNs and QSOs, each of which 
was observed twice in the period of $1996-1998$ in  time interval of 
a year or more separating the observations.  
About a quater of the AGNs in each category are radio loud.  
The AGNs were selected covering a wide range of redshift 
from $z=0$ to 1, and of  absolute 
$B$-magnitude from $M_B=-30$ mag to $-20$ mag.  

The effect of PSF variation cannot be canceled completely in the 
differential photometry, because the target AGNs are not always 
seen as having the same PSF profiles as the surrounding reference 
objects.  The systematic error arising from this PSF effect is 
estimated as about 0.01mag, which is smaller than other errors.  
Considering all systematic errors from different sources, 
we were able to use a small aperture of 3 pixels,  comparable to the 
seeing size.  As a result, smaller statistical error was realized in 
the determination of AGN variability by differential photometry than by 
usual standards-based photometry.  In order to check the accuracy of 
our result, we also applied the differential photometry to the reference 
objects around AGNs, and confirmed that the above error estimation was 
reasonable, independent of the passband.

Significance of AGN variability was measured by a ratio of variability 
relative to estimated error 
$R=\Delta m_{\rm AGN}/\sigma_{\Delta m_{\rm AGN}}$.  Since 
the $R$-values of the AGNs in our sample are similarly distributed 
irrespective of passband, we  average the  $JHK'$ variabiliy 
for the AGNs of $\sigma_{\Delta m_{\rm AGN}}<0.1$ mag.  
Then, the rate of variability 
detection, or the fraction of certainly varied AGNs in this sample is 
58\%($2\sigma$) or 44\%(3$\sigma$).

We furthermore divide the sample into three groups of different error 
ranges such as high accuracy of $\sigma_{\Delta m_{\rm AGN}}<0.03$, 
intermediate accuracy of $\sigma_{\Delta m_{\rm AGN}}=0.03-0.05$, and 
low accuracy of $\sigma_{\Delta m_{\rm AGN}}=0.05-0.1$.  We then 
estimate the rate of variability detection in each of three groups.  
The frequency distribution of the AGNs with respect to $R$ for the 
low-accuracy group was skrewed towards smaller $R$. On the other hand, 
for the high-accuracy group, the frequency distribution shows a 
prominant peak at $R>5$, and the fraction of certainly varied AGNs in 
this group is 80\%($2\sigma$) or 68\%(3$\sigma$).  Thus, the rate of 
variability detection is larger for higher accuracy.  Since this tendency 
shows  no saturation, it does not exclude the 
possibility that all AGNs are indeed variable.

In conclusion, the high detection rate in this paper is mainly obtained 
by our use of differential photometry with a small aperture of 3 pixels.  
The advantage of using differential photometry is to enable an 
efficient detection of variability in as many AGNs as possible 
with  observations made only twice, with a long time interval 
of a year or more separating the observations.  
Furthermore, the advantage of using a small 
aperture is not only to improve the S/N ratio but also to decrease 
the contribution of the AGN host galaxy within the aperture, 
especially for nearby AGNs.

We are grateful to H. Okuda, M. Narita and other staff of the infrared 
astronomy group of  the Institute of Space and Astronautical Science 
(ISAS) for their support in using their 1.3m telescope. We thank the 
staff of the Advanced Technology Center of the National Astronomical 
Observatory of Japan (NAOJ) for their new coating  of the primary mirror 
of the 1.3m telescope at the ISAS.  Gratitude is also extended to the 
Computer Data Analysis Center of the NAOJ.  This work has made use of 
the NASA/IPAC Extra Galactic Database (NED), and has been supported 
partly by the Grand-in-Aid (07CE2002, 10304014) of the Ministry of 
Education, Science, Culture, and Sports of Japan and by the Torey 
Science Foundation.



\clearpage
\figcaption{An example of measuring the AGN variability by differential 
photometry.
        Shown are the differences of two $H$-magnitudes measured on 
96/02/11 and 
        97/12/27 for Mark 1320 ($i=0$) and its nearby reference objects 
($i=1$ to 5). 
        Mark 1320 has brightened by 0.55 mag, relative to the fiducial 
(dashed line) 
        defined by the reference objects.  
      }

\figcaption{The magnitude correction, $\Delta m_{\rm PSF}$, 
           for the effect of PSF variation
           in the estimation of AGN variability, 
           based on the data for which the magnitude 
           accuracy is less than 0.02 mag.
           {\it Top panel:}  
           The relation between the 
           correction, $\Delta m_{\rm PSF}$, and the difference of  
           the FWHMs measured on two  different nights, 
           $\Delta {\rm FWHM}$.  
           {\it Bottom panel:} The histogram of $\Delta m_{\rm PSF}$.
           }

\figcaption{The median of corrected variability, 
            $\Delta m_c$, and its dispersion, 
           $\sigma_{\Delta m_c}({\rm dsp})$, for reference objects 
           as a function of  calculated error, 
           $\sigma_{\Delta m_c}({\rm cal})$.   
           Triangles, squares, and circles represent the $J$, 
           $H$, and $K'$ bands, repectively.
          The filled symbols are for $\Delta m_c$, and the open symbols 
          are for $\sigma_{\Delta m_c}({\rm dsp})$.
         }

\figcaption{A comparison between AGN variabilities determined by 
            differential photometry, $\Delta m_{\rm AGN}$(dif), and 
            standards-based photometry, $\Delta m_{\rm AGN}$(std).
       Shown are the $J$-band results for various apertures 
       of 7, 10, 12, and 15  pixels.
        }

\figcaption{A comparison between errors of AGN variabilities determined by 
            differential photometry, $\sigma_{\Delta m_{\rm AGN}}$(dif), 
            and standards-based photometry, $\sigma_{\Delta m_{\rm 
AGN}}$(std).
            Shown are the $J$-band results for various apertures 
            of 7, 10, 12, and 15  pixels.
        }

\figcaption{
            The fractional number of AGNs, $f(\ge c)$, as a function of  
            aperture for various values of error ratio defined by 
             $c=\sigma_{\Delta m_{\rm AGN}}({\rm std})/
            \sigma_{\Delta m_{\rm AGN}}({\rm dif})$.  The top, middle, 
            and bottom panels are for  the $J$, $H$, and $K'$ bands, 
            respectively.
        }

\figcaption{The distribution of AGN variability, $\Delta m_{\rm AGN}$, and 
            redshift $z$, based on the data with more than two 
            reference objects and with an accuracy higher than 0.1 mag.
            The top, middle, and bottom panels are for  the $J$, $H$, and 
$K'$ 
            bands, respectively.
         }

\figcaption{The frequency distribution of variability, $\Delta m_{\rm AGN}$,
            based on the data with more than two reference objects and with 
            an accuracy higher than 0.1 mag.   
           }

\figcaption{The frequency distribution of AGNs having varied with respect 
to the
         ratio of $R=\Delta m_{\rm AGN}/\sigma_{\Delta m_{\rm AGN}}$.  
         The top, middle, and bottom panels are for all AGNs, 
         radio-quiet AGNs, and radio-loud AGNs, respectively.
         }

\figcaption{The requency distribution of AGNs 
        having varied with respect to the
        ratio of $R=\Delta m_{\rm AGN}/\sigma_{\Delta m_{\rm AGN}}$.  
        Same as in Fig. 9, but for radio-quiet AGNs (left column) and 
        radio-loud AGNs (right column) subdivided by the rest-frame time 
        interval, $\Delta t_{\rm rest}$, between observations.
          }

\figcaption{The frequency distribution of AGNs having 
         varied with respect to the
          ratio of $R=\Delta m_{\rm AGN}/\sigma_{\Delta m_{\rm AGN}}$.  
         Same as in Fig. 9, but for all AGNs subdivided by their 
          variability error, $\sigma_{\Delta m_{\rm AGN}}$. 
        }


\clearpage
\begin{figure}
\begin{center}
  Figure 1.
  \plotone{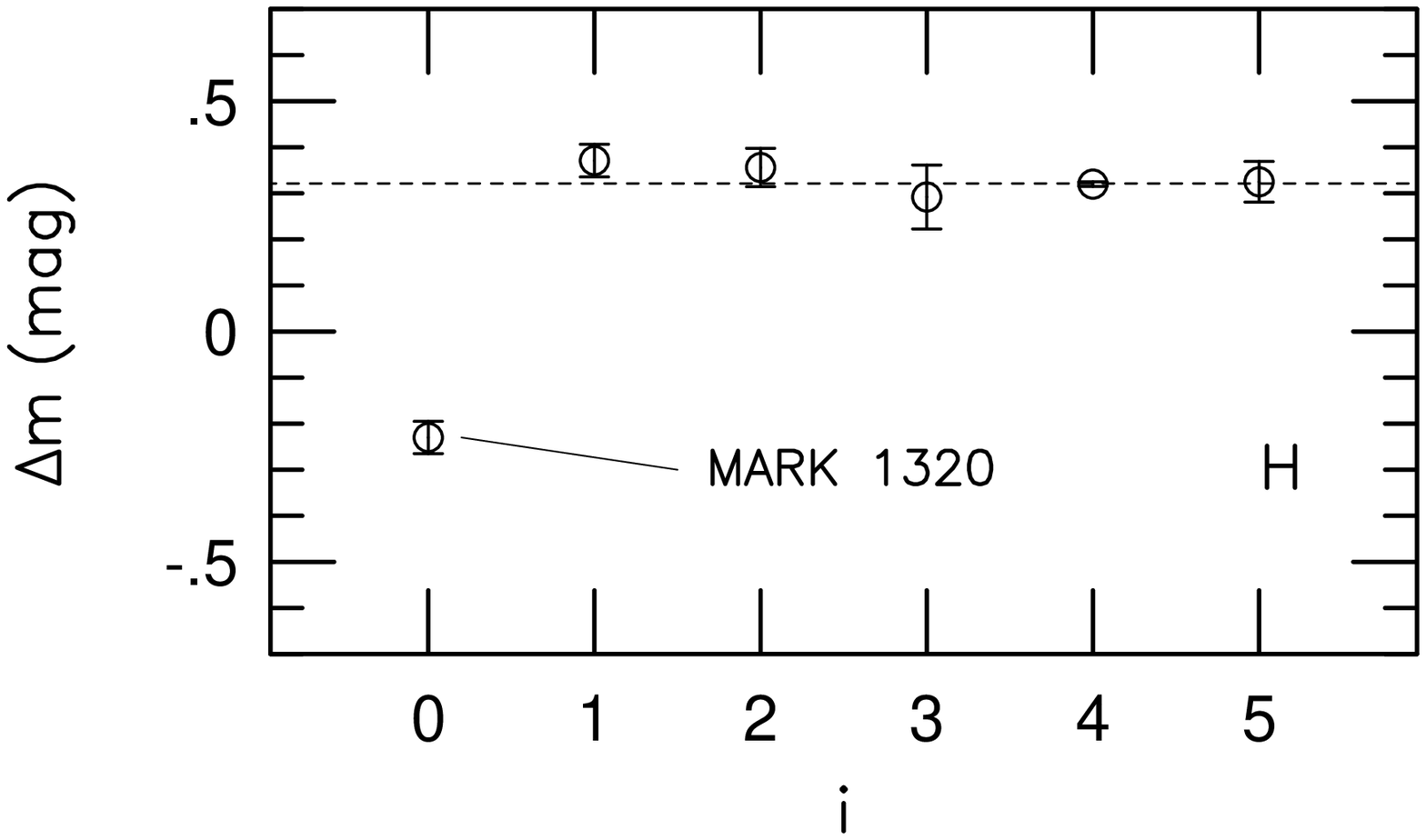}
\end{center}
\end{figure}

\clearpage
\begin{figure}
\begin{center}
  Figure 2.
  \epsscale{0.9} 
  \plotone{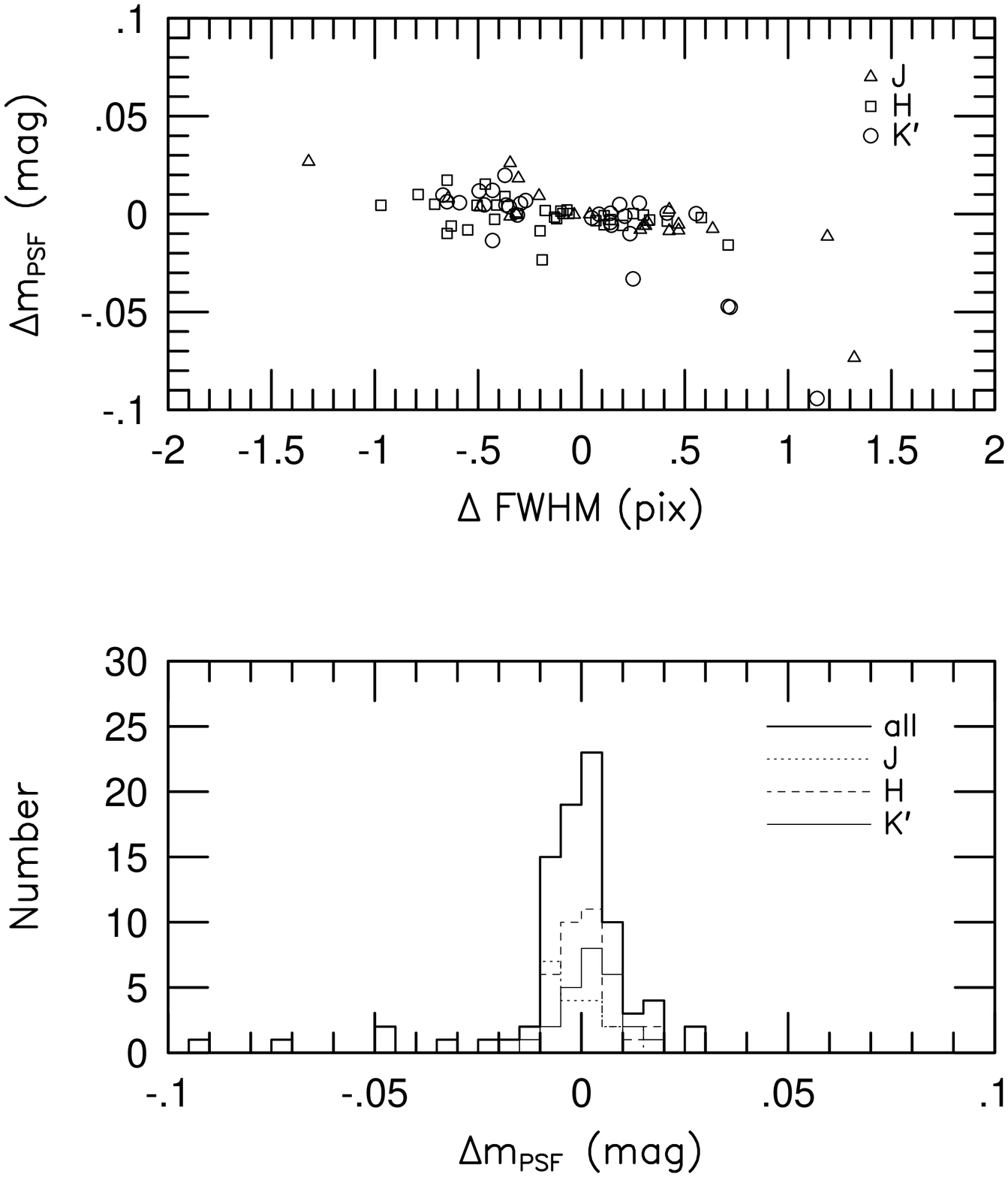}
\end{center}
\end{figure}

\clearpage
\begin{figure}
\begin{center}
  Figure 3.
  \epsscale{0.9} 
  \plotone{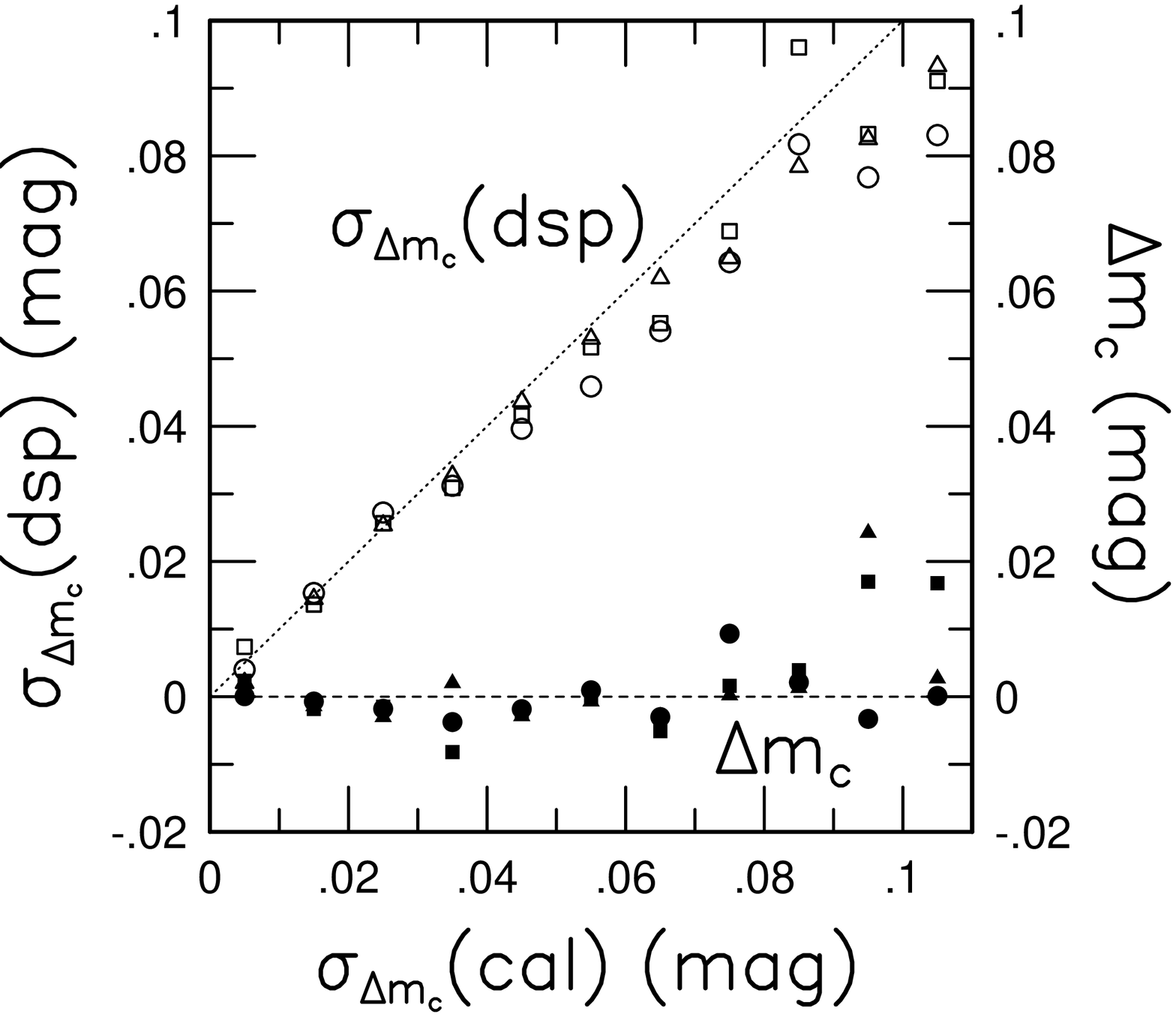}
\end{center}
\end{figure}

\clearpage
\begin{figure}
\begin{center}
  Figure 4.
  \plotone{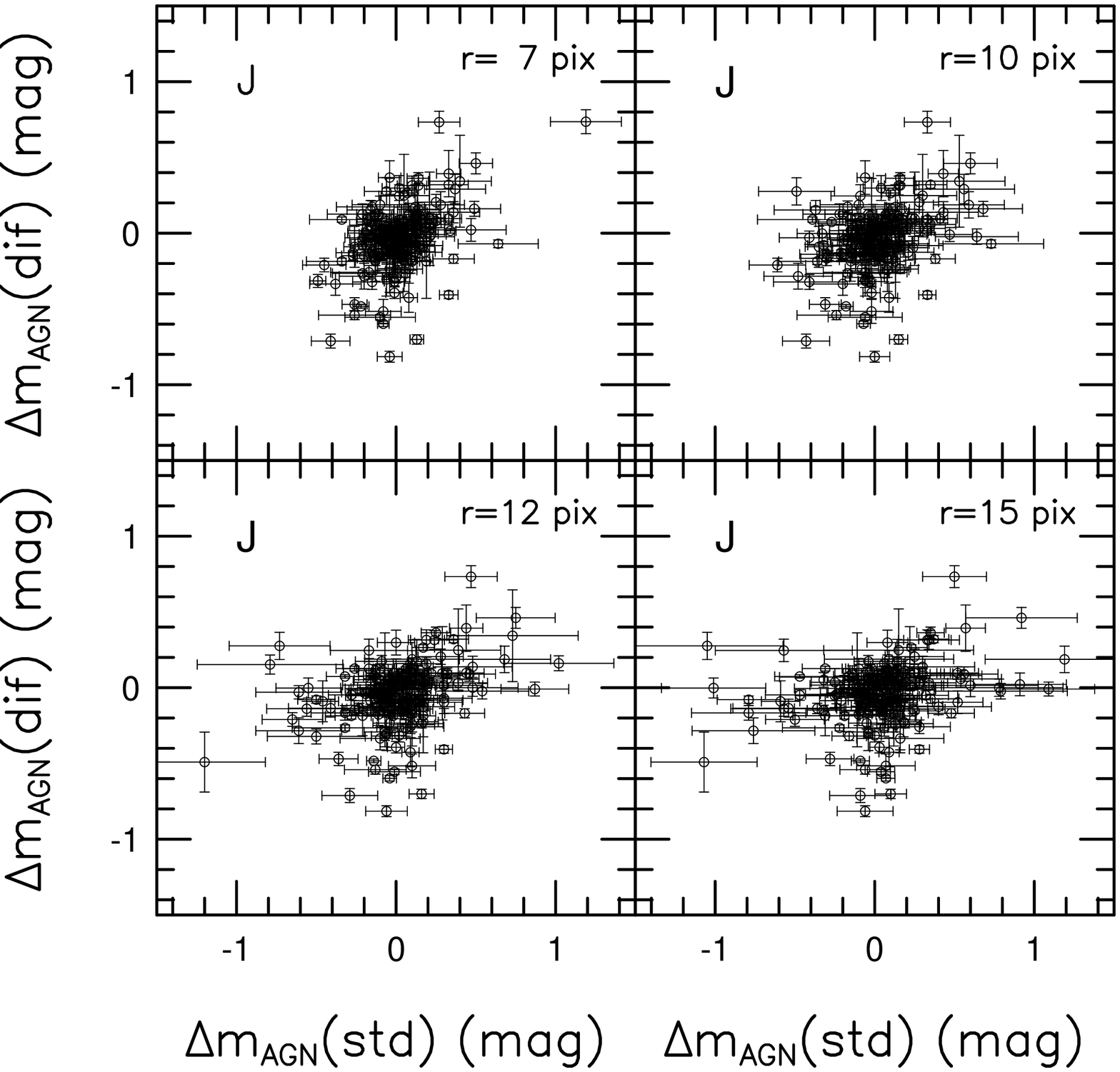}
\end{center}
\end{figure}

\clearpage
\begin{figure}
\begin{center}
  Figure 5.
  \plotone{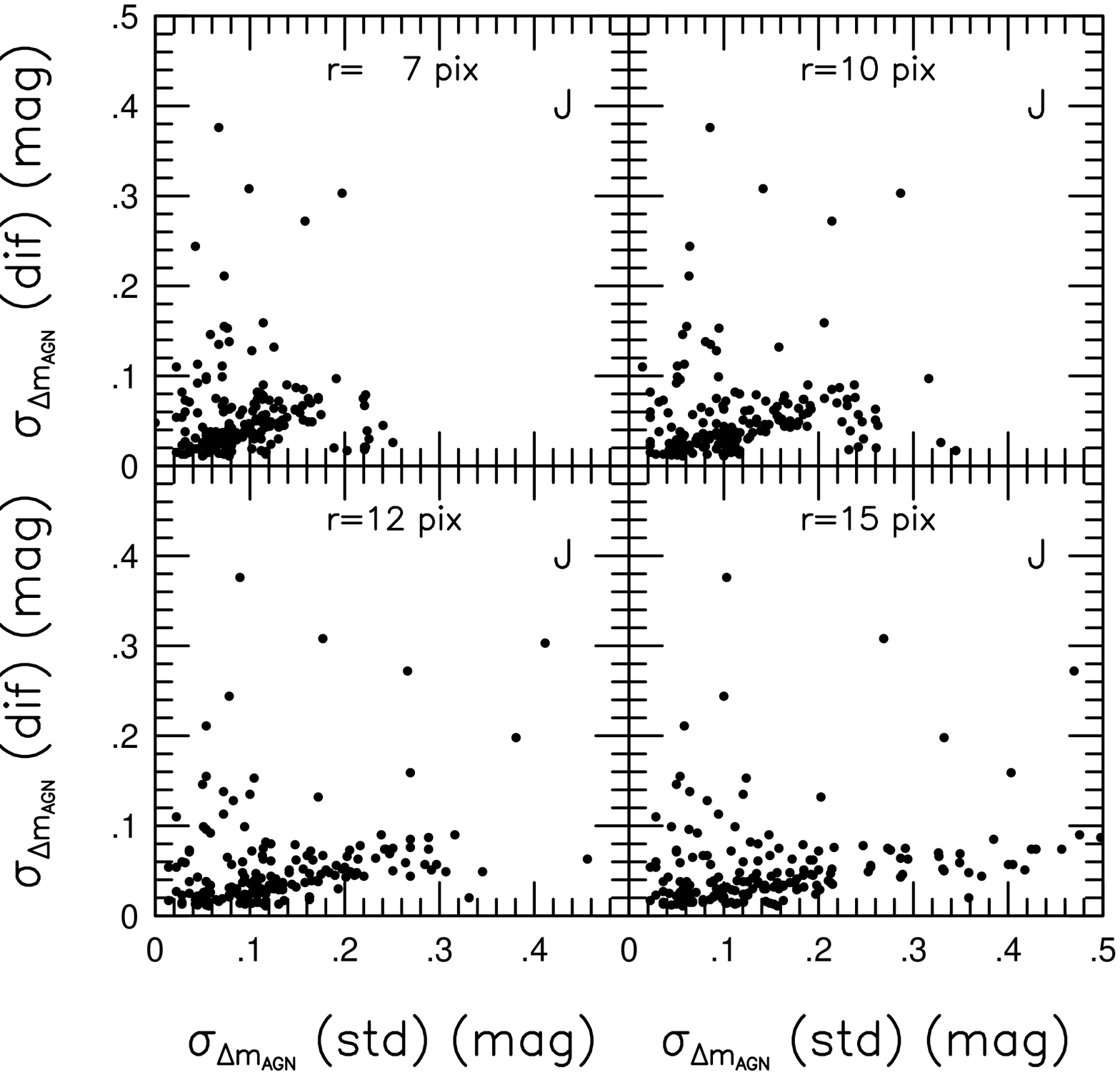}
\end{center}
\end{figure}

\clearpage
\begin{figure}
\begin{center}
  Figure 6.\\
  \epsscale{0.8}
  \plotone{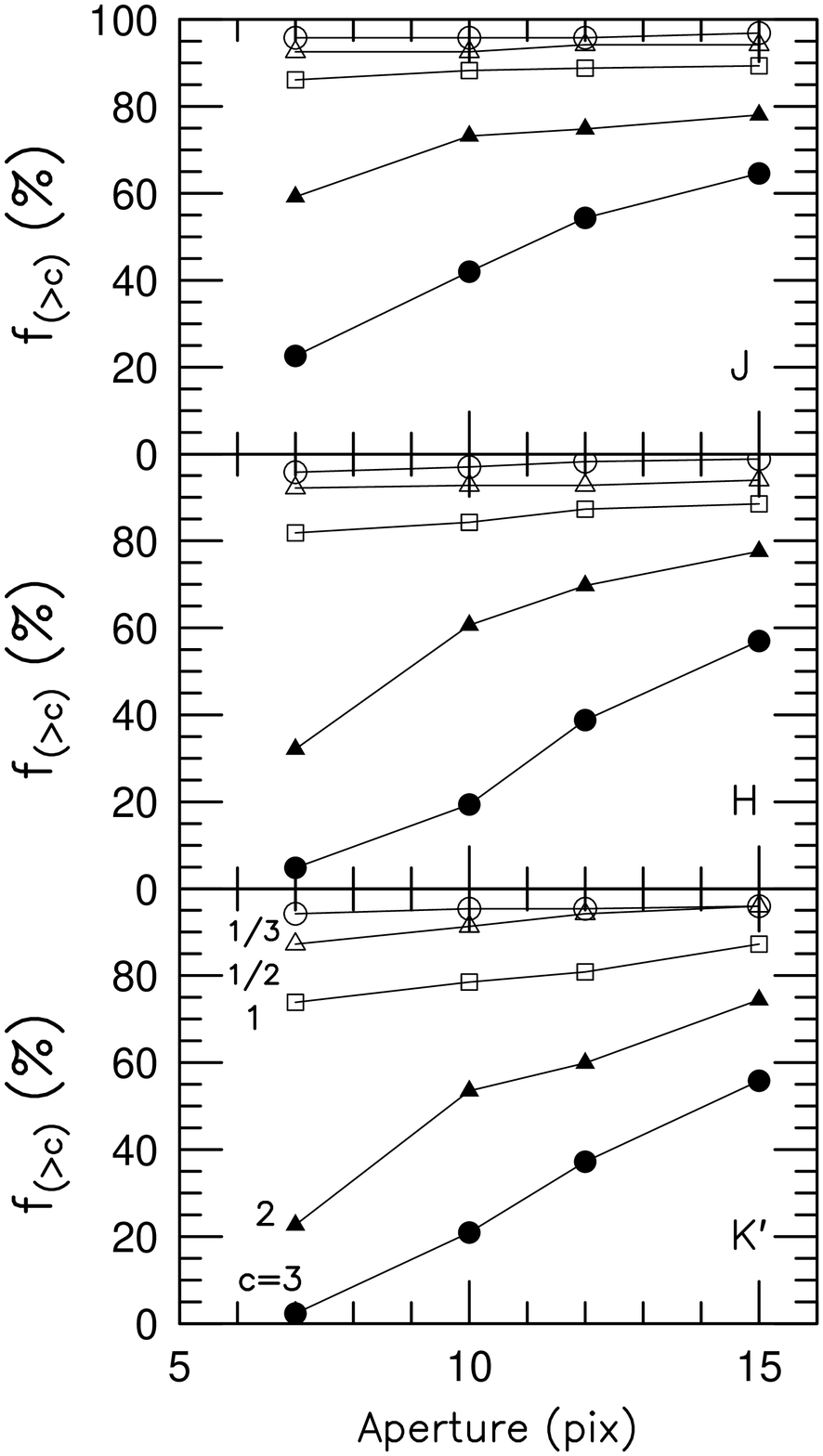}
\end{center}
\end{figure}

\clearpage
\begin{figure}
\begin{center}
  Figure 7.\\
  \epsscale{0.75}
  \plotone{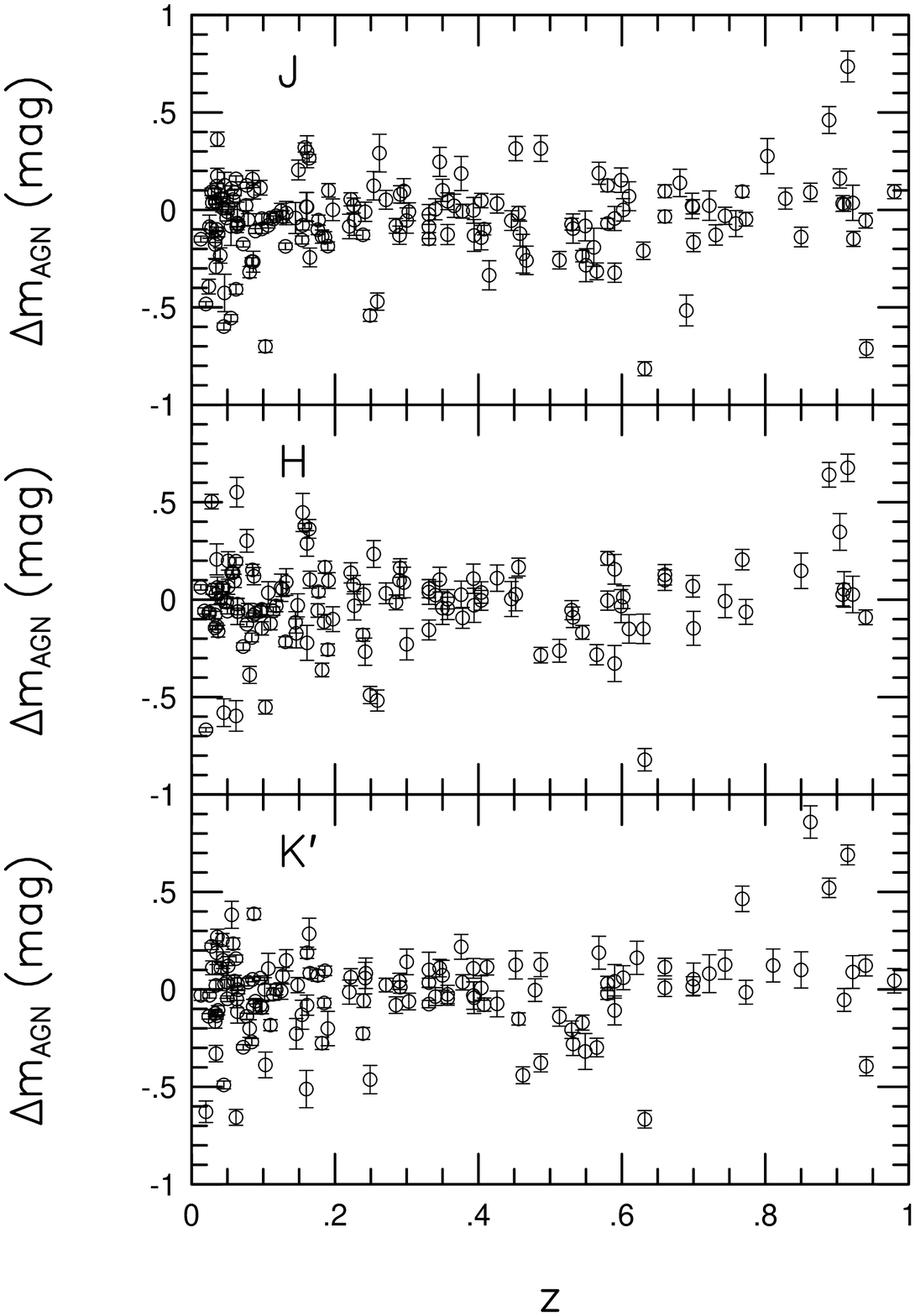}
\end{center}
\end{figure}

\clearpage
\begin{figure}
\begin{center}
  Figure 8.
  \epsscale{1}
  \plotone{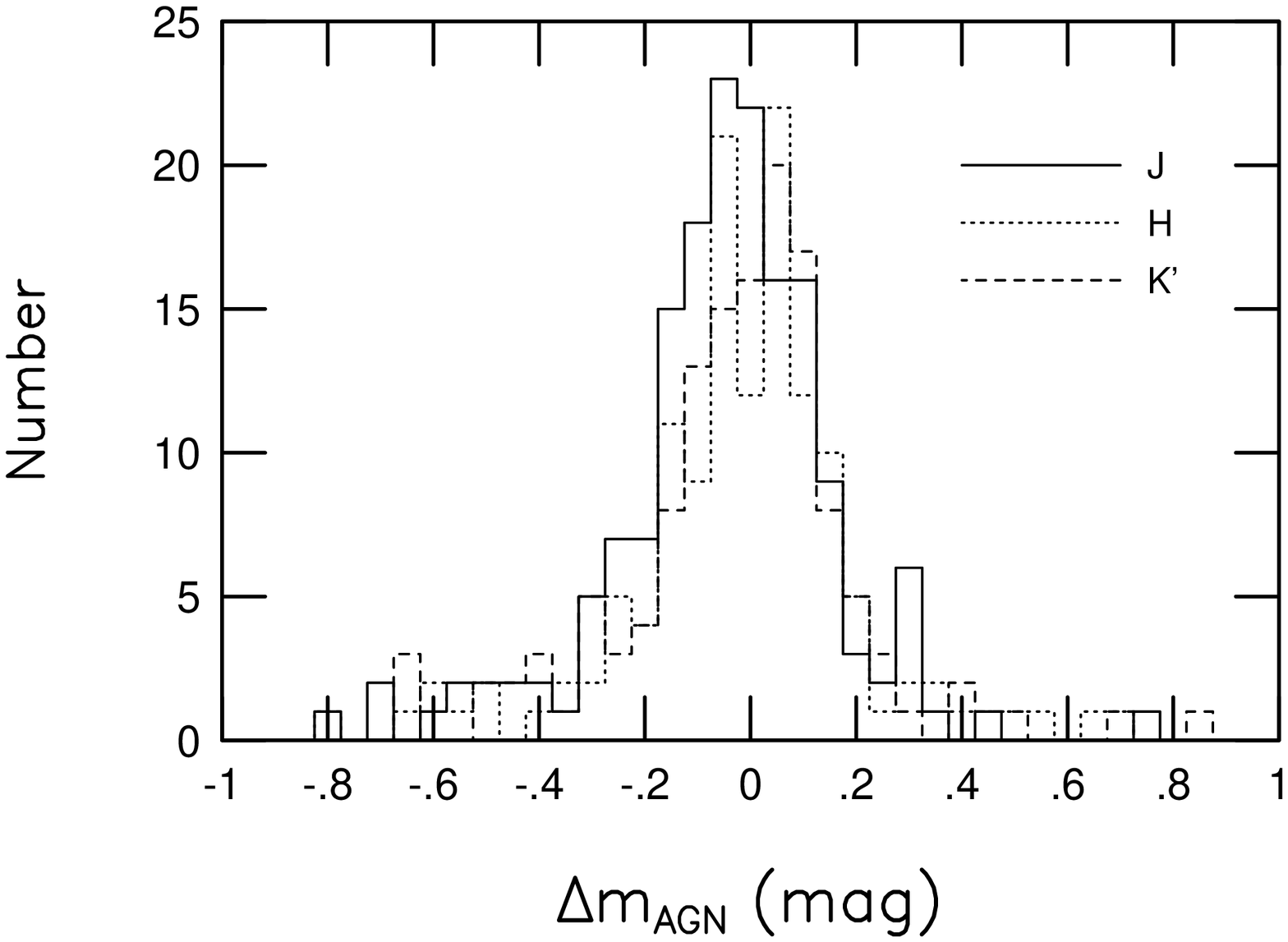}
\end{center}
\end{figure}

\clearpage
\begin{figure}
\begin{center}
  Figure 9.
  \epsscale{1}
  \plotone{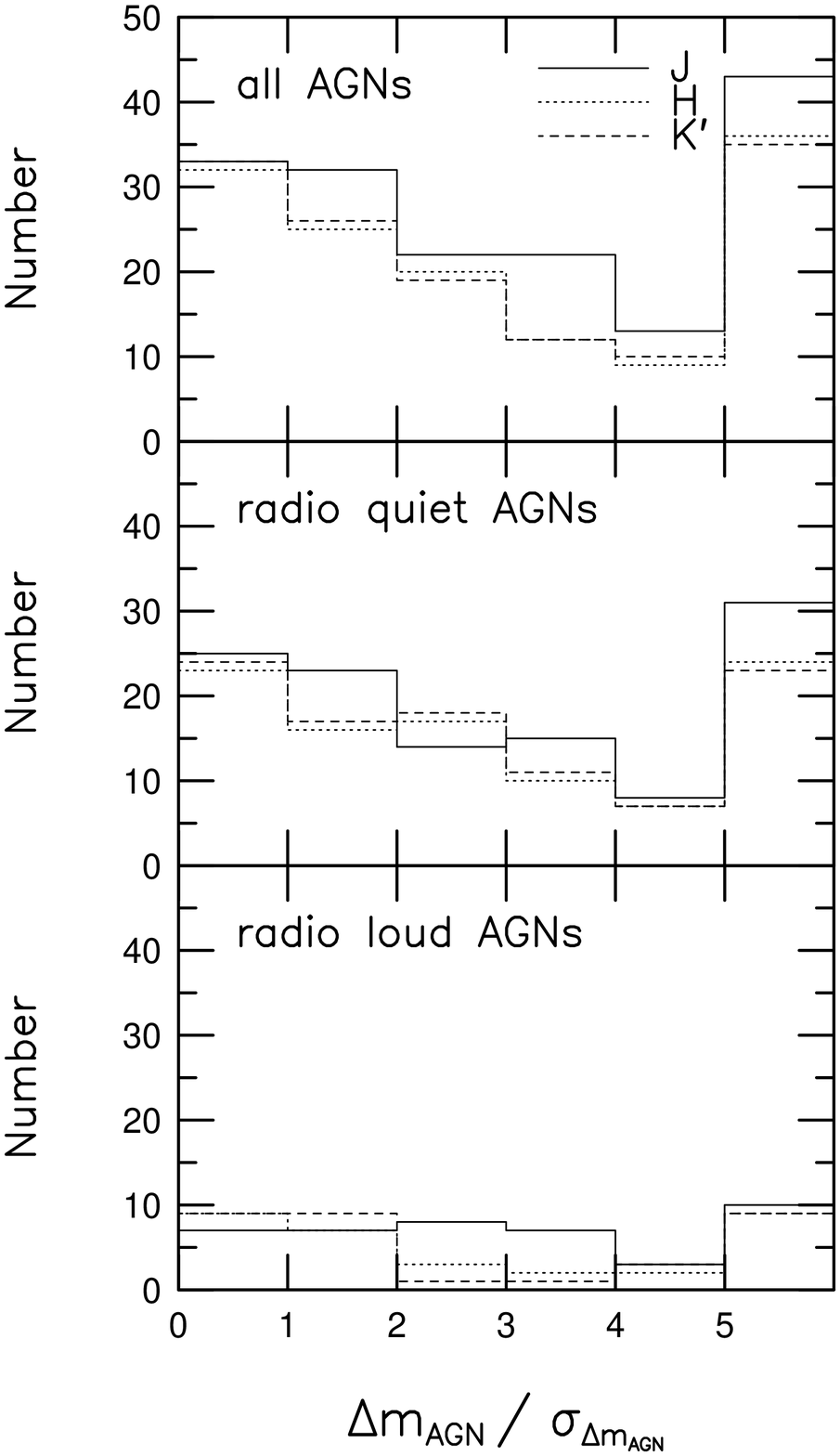}
\end{center}
\end{figure}

\clearpage
\begin{figure}
\begin{center}
  Figure 10.
  \epsscale{1}
  \plotone{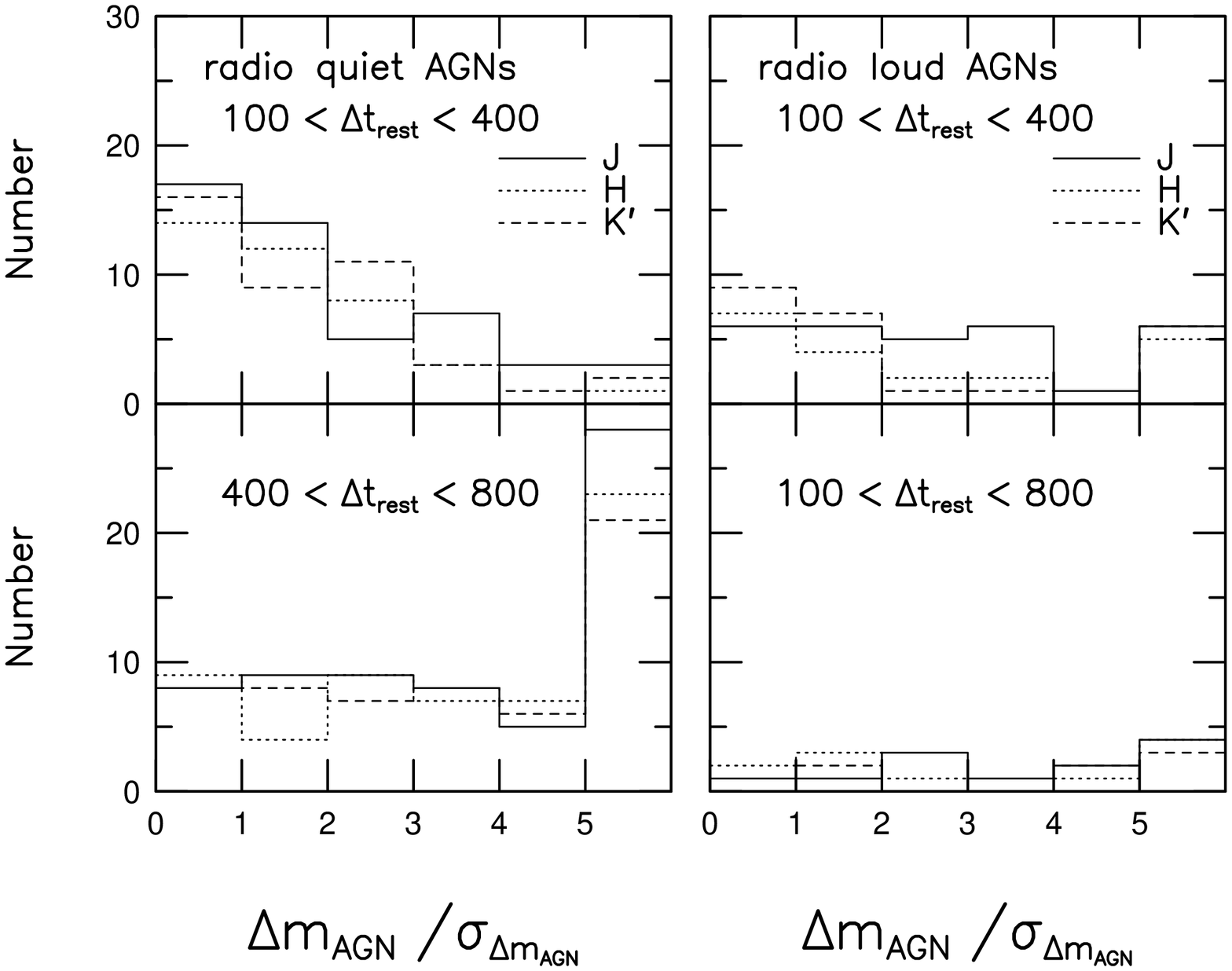}
\end{center}
\end{figure}

\clearpage
\begin{figure}
\begin{center}
  Figure 11.
  \epsscale{1}
  \plotone{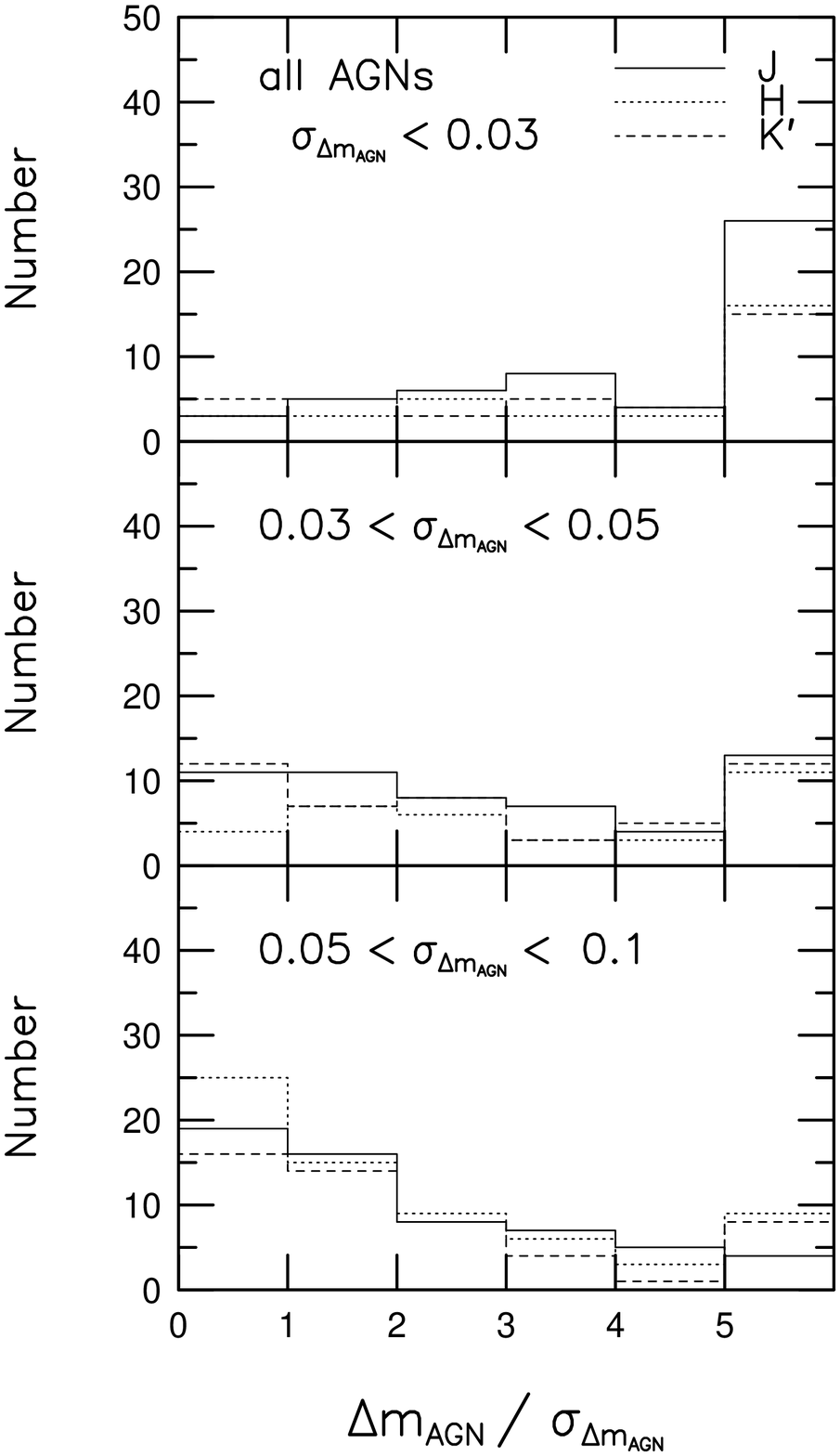}
\end{center}
\end{figure}


\clearpage

\begin{deluxetable}{clccccccccccc}
\scriptsize
\tablenum{1}
\tablecaption{Variability data of 226 AGNs}
\tablewidth{0pt}
\tablehead{
\colhead{Number} & \colhead{Name}   
& \colhead{$\Delta J$}   & \colhead{$\sigma_{\Delta J}$}  & \colhead{$n_J$}  
& \colhead{$\Delta H$}   & \colhead{$\sigma_{\Delta H}$}  & \colhead{$n_H$}   
& \colhead{$\Delta K'$}  & \colhead{$\sigma_{\Delta K'}$} & \colhead{$n_{K'}$}   
& \colhead{date1}   & \colhead{date2}
} 
\startdata
   1  &      PB 5669           &   $-$    &   $-$    &   $-$  &   0.366  &   0.126  &     6  &  -0.004  &   0.058  &     6  &    96/12/03   &   98/01/05 \nl
   5  &      PB 5853           &   0.017  &   0.073  &     5  &   $-$    &   $-$    &   $-$  &  -0.511  &   0.096  &     8  &    96/12/03   &   98/01/21 \nl
   7  &      PB 5932           &   0.048  &   0.037  &    10  &   0.037  &   0.055  &     5  &   0.162  &   0.105  &     9  &    96/12/01   &   98/01/26 \nl
   8  &      MS 00377-0156     &   0.097  &   0.063  &     2  &   0.088  &   0.078  &     8  &   $-$    &   $-$    &   $-$  &    96/12/03   &   98/01/06 \nl
  14  &      PKS 0130+24       &   0.315  &   0.062  &    11  &   0.028  &   0.085  &    10  &   0.126  &   0.072  &     9  &    96/12/01   &   98/01/21 \nl
  15  &      UM 341            &  -0.030  &   0.082  &     1  &   $-$    &   $-$    &   $-$  &   0.031  &   0.076  &     1  &    96/12/01   &   97/12/27 \nl
  19  &      KUV 01507-0744    &  -0.054  &   0.065  &     3  &  -0.229  &   0.081  &     3  &   0.141  &   0.066  &     2  &    96/11/25   &   97/12/31 \nl
  20  &      PHL 1226          &  -0.142  &   0.063  &     5  &  -0.004  &   0.051  &     4  &   0.007  &   0.063  &     6  &    96/12/01   &   98/01/16 \nl
  21  &      UM 381            &   0.027  &   0.034  &     3  &   0.053  &   0.091  &     3  &  -0.054  &   0.059  &     4  &    96/12/08   &   98/01/04 \nl
  22  &      UM 153            &   0.096  &   0.033  &     9  &   0.100  &   0.053  &     9  &   0.115  &   0.045  &     7  &    96/12/08   &   98/01/05 \nl
  23  &      MARK 1018         &   0.011  &   0.022  &     7  &   0.061  &   0.019  &     5  &   0.254  &   0.033  &     2  &    96/02/11   &   98/01/05 \nl
  24  &      RXS J02070+2930   &  -0.041  &   0.023  &     7  &  -0.122  &   0.033  &     5  &  -0.183  &   0.029  &     5  &    96/02/11   &   98/01/05 \nl
  25  &      MARK  586         &  -0.079  &   0.024  &     8  &   0.448  &   0.098  &     6  &   0.000  &   0.204  &     6  &    96/02/07   &   98/01/05 \nl
  26  &      PKS 0214+10       &  -0.098  &   0.030  &     6  &  -0.072  &   0.108  &     8  &  -0.078  &   0.034  &     7  &    96/12/08   &   97/12/27 \nl
  27  &      PB 9130           &   0.544  &   0.100  &     1  &   0.126  &   0.102  &     2  &   0.062  &   0.106  &     1  &    96/11/25   &   96/12/01 \nl
  28  &      B3 0219+443       &  -0.139  &   0.050  &    26  &   0.148  &   0.091  &    21  &   0.100  &   0.093  &    13  &    96/12/22   &   98/01/31 \nl
  32  &      MS 02328-0400     &   0.187  &   0.087  &     4  &   0.025  &   0.070  &     6  &   0.218  &   0.064  &     4  &    96/12/21   &   98/01/16 \nl
  34  &      Q 0235+0121       &  -0.002  &   0.066  &     5  &   0.108  &   0.075  &     5  &  -0.032  &   0.070  &     3  &    96/12/02   &   98/01/04 \nl
  35  &      Q 0238-0142       &   0.246  &   0.074  &     5  &   0.103  &   0.064  &     5  &   0.109  &   0.045  &     5  &    96/12/08   &   98/02/17 \nl
  37  &      US 3150           &  -0.259  &   0.073  &     5  &   $-$    &   $-$    &   $-$  &    $-$   &    $-$   &   $-$  &    96/11/25   &   98/01/06 \nl
  38  &      MS 02448+1928     &  -0.101  &   0.025  &     8  &  -0.056  &   0.035  &     6  &   0.071  &   0.029  &     7  &    96/12/01   &   97/12/27 \nl
  42  &      US 3376           &  -0.114  &   0.129  &     1  &   $-$    &   $-$    &   $-$  &    $-$   &    $-$   &   $-$  &    96/12/22   &   97/02/22 \nl
  44  &      US 3472           &  -0.096  &   0.075  &     5  &  -0.090  &   0.053  &     4  &  -0.280  &   0.059  &     6  &    96/12/22   &   97/12/31 \nl
  45  &      S 0257-0027       &   0.070  &   0.142  &     7  &  -0.336  &   0.192  &     7  &   0.000  &   0.060  &     6  &    97/02/09   &   98/01/21 \nl
  50  &      Q 0305+0222       &  -0.044  &   0.062  &     6  &   0.156  &   0.076  &     6  &  -0.108  &   0.074  &     5  &    96/12/02   &   98/02/09 \nl
  51  &      PKS 0306+102      &   $-$    &   $-$    &   $-$  &    $-$   &    $-$   &   $-$  &   0.859  &   0.083  &     5  &    96/12/23   &   98/01/26 \nl
  54  &      MS 03120+1405     &  -0.030  &   0.044  &     8  &  -0.007  &   0.085  &     7  &   0.127  &   0.075  &     7  &    96/12/09   &   98/01/29 \nl
  56  &      B2 0321+33        &   0.159  &   0.015  &    24  &   0.197  &   0.019  &    19  &   0.158  &   0.018  &    12  &    96/02/05   &   98/01/05 \nl
  59  &      3C  93.0          &  -0.126  &   0.052  &     9  &  -0.045  &   0.058  &    10  &  -0.025  &   0.047  &    10  &    97/02/04   &   98/01/05 \nl
  61  &      PKS 0353+027      &   0.000  &   0.057  &    11  &   0.013  &   0.059  &    16  &   0.060  &   0.062  &    13  &    97/01/22   &   98/01/16 \nl
  62  &      MS 03574+1046     &  -0.137  &   0.034  &     4  &  -0.361  &   0.035  &     7  &  -0.276  &   0.032  &     6  &    96/12/01   &   97/12/24 \nl
  64  &      MS 04124-0802     &  -0.123  &   0.015  &     6  &  -0.162  &   0.030  &     7  &  -0.107  &   0.024  &     5  &    96/02/06   &   97/12/27 \nl
  65  &      3C 110            &  -0.048  &   0.035  &     8  &  -0.063  &   0.064  &     5  &  -0.014  &   0.061  &     7  &    96/02/02   &   97/12/31 \nl
\enddata
\end{deluxetable}

\begin{deluxetable}{clccccccccccc}
\scriptsize
\tablenum{1-Continued}
\tablecaption{}
\tablewidth{0pt}
\tablehead{
\colhead{Number} & \colhead{Name}   
& \colhead{$\Delta J$}   & \colhead{$\sigma_{\Delta J}$}  & \colhead{$n_J$}  
& \colhead{$\Delta H$}   & \colhead{$\sigma_{\Delta H}$}  & \colhead{$n_H$}   
& \colhead{$\Delta K'$}  & \colhead{$\sigma_{\Delta K'}$} & \colhead{$n_{K'}$}   
& \colhead{date1}   & \colhead{date2}
} 
\startdata
  66  &      PKS 0420-01       &   0.736  &   0.079  &     7  &   0.677  &   0.070  &     6  &   0.690  &   0.051  &     4  &    96/12/08   &   97/12/24 \nl
  67  &      3C 120            &   0.044  &   0.014  &    14  &  -0.079  &   0.018  &    11  &  -0.165  &   0.034  &     6  &    96/02/05   &   97/12/27 \nl
  68  &      IRAS 04448-0513   &   0.053  &   0.014  &     9  &   0.060  &   0.021  &     8  &   0.150  &   0.043  &     6  &    96/02/05   &   97/12/27 \nl
  69  &      Q 0446+0130       &   $-$    &   $-$    &   $-$  &    $-$   &    $-$   &   $-$  &   0.123  &   0.085  &     8  &    97/01/20   &   98/01/31 \nl
  71  &      UGC  3223         &  -0.305  &   0.110  &     3  &  -0.058  &   0.016  &     9  &   $-$    &   $-$    &   $-$  &    96/02/06   &   97/12/27 \nl
  72  &      2E 0507+1626      &   0.007  &   0.083  &     1  &   $-$    &   $-$    &   $-$  &   0.468  &   0.152  &     1  &    96/02/06   &   97/12/27 \nl
  73  &      3C 135.0          &  -0.035  &   0.030  &    18  &   0.053  &   0.037  &    17  &   0.070  &   0.042  &    16  &    96/12/07   &   97/12/31 \nl
  74  &      AKN  120          &  -0.177  &   0.072  &    22  &  -0.145  &   0.011  &    16  &  -0.124  &   0.014  &    11  &    96/02/05   &   98/01/06 \nl
  75  &      1E 0514-0030      &   0.080  &   0.044  &    19  &   0.163  &   0.048  &    14  &   0.011  &   0.031  &    18  &    96/02/11   &   97/12/24 \nl
  76  &      3C 138.0          &  -0.070  &   0.067  &    48  &   $-$    &   $-$    &   $-$  &    $-$   &    $-$   &   $-$  &    97/01/06   &   98/01/24 \nl
  77  &      3C 147.0          &  -0.237  &   0.029  &    24  &  -0.168  &   0.036  &    25  &  -0.171  &   0.039  &    18  &    96/12/21   &   98/01/05 \nl
  78  &      4C 16.14          &  -0.491  &   0.198  &    70  &   $-$    &   $-$    &   $-$  &    $-$   &    $-$   &   $-$  &    96/12/07   &   98/02/08 \nl
  79  &      MCG  08.11.11     &  -0.483  &   0.012  &    59  &  -0.668  &   0.011  &    24  &  -0.628  &   0.055  &    16  &    96/02/03   &   98/01/06 \nl
  81  &      3C 154.0          &   0.126  &   0.032  &    62  &   0.212  &   0.035  &    61  &  -0.022  &   0.030  &    50  &    96/12/09   &   97/12/31 \nl
  82  &      MC 0657+176       &   0.022  &   0.075  &    57  &   $-$    &   $-$    &   $-$  &   0.081  &   0.097  &    40  &    97/01/06   &   98/02/06 \nl
  83  &      3C 175.0          &   0.094  &   0.029  &     3  &   0.205  &   0.052  &    12  &   0.465  &   0.065  &    10  &    96/11/29   &   98/01/04 \nl
  84  &      B2 0709+37        &   0.315  &   0.067  &    17  &  -0.284  &   0.040  &    16  &  -0.377  &   0.046  &    20  &    96/02/02   &   98/01/05 \nl
  85  &      MARK  376         &   0.112  &   0.011  &    33  &   0.140  &   0.012  &    25  &   0.383  &   0.069  &    15  &    96/02/03   &   98/01/05 \nl
  88  &      PKS 0736+01       &   0.100  &   0.035  &    20  &   0.098  &   0.040  &    20  &  -2.699  &   0.111  &     1  &    96/01/31   &   97/12/18 \nl
  89  &      OI 363            &  -0.210  &   0.045  &    19  &  -0.150  &   0.076  &    18  &   $-$    &   $-$    &   $-$  &    96/01/31   &   98/01/06 \nl
  91  &      B2 0742+31        &  -0.225  &   0.099  &    18  &  -0.200  &   0.102  &    18  &  -0.441  &   0.043  &    13  &    96/01/24   &   97/12/24 \nl
  93  &      GC 0742+33        &   0.070  &   0.074  &    19  &  -0.150  &   0.073  &    19  &   $-$    &   $-$    &   $-$  &    96/12/21   &   98/01/21 \nl
  95  &      RXS J07498+3454   &  -0.016  &   0.060  &     9  &   0.090  &   0.069  &     5  &   0.149  &   0.055  &     6  &    96/12/03   &   98/01/05 \nl
  96  &      PKS 0748+126      &   0.461  &   0.069  &    13  &   0.641  &   0.063  &    13  &   0.521  &   0.050  &    16  &    97/01/26   &   98/01/16 \nl
  98  &      B2 0752+25A       &  -0.055  &   0.051  &     6  &   0.003  &   0.090  &     3  &  -0.182  &   0.124  &     2  &    96/01/31   &   97/12/31 \nl
  99  &      B3 0754+394       &   0.173  &   0.146  &    17  &   0.126  &   0.107  &     5  &   0.058  &   0.011  &    11  &    96/01/24   &   98/01/05 \nl
 101  &      UGC  4155         &  -0.086  &   0.059  &    14  &  -0.068  &   0.013  &    12  &  -0.030  &   0.014  &     6  &    96/02/06   &   97/12/24 \nl
 102  &      MARK 1210         &  -0.150  &   0.015  &    25  &   0.063  &   0.014  &    17  &  -0.030  &   0.026  &    15  &    96/02/05   &   98/01/05 \nl
 103  &      MS 08019+2129     &  -0.030  &   0.022  &    12  &  -0.032  &   0.030  &     6  &   0.000  &   0.027  &     6  &    96/01/24   &   97/12/18 \nl
 104  &      3C 192.0          &  -0.144  &   0.138  &     3  &  -0.118  &   0.139  &     2  &   0.020  &   0.087  &     1  &    96/12/03   &   97/12/18 \nl
 105  &      MS 08080+4840     &  -0.166  &   0.048  &    16  &  -0.147  &   0.087  &    14  &   0.052  &   0.083  &    16  &    96/12/31   &   98/02/10 \nl
 109  &      RX J08166+2941    &   0.291  &   0.097  &    15  &   $-$    &   $-$    &   $-$  &   0.604  &   0.107  &    15  &    96/12/07   &   98/01/19 \nl
 110  &      3C 197            &   0.094  &   0.036  &    20  &   0.072  &   0.105  &    17  &   0.044  &   0.063  &    16  &    97/01/26   &   98/01/26 \nl
\enddata
\end{deluxetable}

\begin{deluxetable}{clccccccccccc}
\scriptsize
\tablenum{1-Continued}
\tablecaption{}
\tablewidth{0pt}
\tablehead{
\colhead{Number} & \colhead{Name}   
& \colhead{$\Delta J$}   & \colhead{$\sigma_{\Delta J}$}  & \colhead{$n_J$}  
& \colhead{$\Delta H$}   & \colhead{$\sigma_{\Delta H}$}  & \colhead{$n_H$}   
& \colhead{$\Delta K'$}  & \colhead{$\sigma_{\Delta K'}$} & \colhead{$n_{K'}$}   
& \colhead{date1}   & \colhead{date2}
} 
\startdata
 111  &      RXS J08223+3305   &  -0.005  &   0.038  &     4  &   0.061  &   0.070  &     4  &  -0.008  &   0.042  &     6  &    97/02/20   &   98/01/16 \nl
 112  &      KUV 08217+4235    &   $-$    &   $-$    &   $-$  &    $-$   &    $-$   &   $-$  &  -0.037  &   0.127  &     2  &    96/11/29   &   97/12/31 \nl
 113  &      4C 44.17          &   0.161  &   0.049  &    12  &   0.347  &   0.094  &    14  &   0.377  &   0.428  &    12  &    97/01/27   &   98/02/09 \nl
 115  &      B2 0827+24        &  -0.712  &   0.046  &    14  &  -0.302  &   0.111  &    13  &  -0.394  &   0.049  &    14  &    96/12/07   &   98/01/16 \nl
 116  &      PG 0832+251       &  -0.148  &   0.032  &     9  &  -0.155  &   0.051  &     8  &  -0.075  &   0.019  &     9  &    96/01/24   &   97/12/24 \nl
 118  &      US 1329           &  -0.541  &   0.030  &    10  &  -0.489  &   0.044  &     6  &  -0.463  &   0.073  &     7  &    96/02/02   &   97/12/25 \nl
 119  &      MARK 1218         &  -0.094  &   0.028  &     2  &  -0.456  &   0.128  &     3  &  -0.179  &   0.139  &     3  &    96/02/05   &   97/12/24 \nl
 122  &      KUV 08377+4136    &  -0.516  &   0.079  &    10  &   0.573  &   0.148  &     9  &   $-$    &   $-$    &   $-$  &    96/12/22   &   98/02/08 \nl
 123  &      PG 0844+349       &  -0.067  &   0.018  &     9  &  -0.023  &   0.030  &     6  &  -0.115  &   0.059  &     3  &    96/02/02   &   97/12/24 \nl
 124  &      55W 179           &   $-$    &   $-$    &   $-$  &    $-$   &    $-$   &   $-$  &  -0.196  &   0.179  &     2  &    97/02/12   &   98/03/18 \nl
 125  &      CSO   2           &  -0.034  &   0.032  &     6  &   0.128  &   0.052  &     9  &   0.008  &   0.044  &     8  &    96/12/07   &   98/01/06 \nl
 127  &      LB 8741           &   0.188  &   0.057  &    15  &   $-$    &   $-$    &   $-$  &   0.188  &   0.084  &     8  &    96/12/07   &   98/01/24 \nl
 129  &      US 1786           &   $-$    &   $-$    &   $-$  &   0.118  &   0.111  &     1  &   0.127  &   0.061  &     8  &    96/12/22   &   98/01/21 \nl
 131  &      MS 08498+2820     &   0.001  &   0.056  &     7  &  -0.100  &   0.064  &    11  &  -0.068  &   0.294  &     4  &    97/02/19   &   98/01/30 \nl
 132  &      MS 08502+2825     &   0.036  &   0.090  &    12  &   0.026  &   0.094  &    12  &   0.089  &   0.084  &    10  &    96/12/02   &   98/01/29 \nl
 133  &      US 1867           &  -0.258  &   0.045  &     7  &  -0.262  &   0.059  &     8  &  -0.141  &   0.048  &     4  &    96/02/07   &   97/12/25 \nl
 135  &      NGC 2683 U1       &   0.343  &   0.303  &     3  &  -0.149  &   0.139  &     4  &   0.162  &   0.085  &     4  &    96/12/08   &   98/01/19 \nl
 136  &      LB 8948           &  -0.023  &   0.051  &     7  &   0.057  &   0.047  &     8  &   0.035  &   0.023  &     6  &    96/12/07   &   97/12/24 \nl
 137  &      LB 8960           &   0.058  &   0.054  &     7  &   $-$    &   $-$    &   $-$  &   0.294  &   0.109  &     7  &    96/12/07   &   98/01/30 \nl
 138  &      US 2068           &  -0.149  &   0.039  &     4  &   $-$    &   $-$    &   $-$  &   0.006  &   0.127  &     4  &    96/11/29   &   98/01/21 \nl
 139  &      KUV 09012+4019    &  -0.005  &   0.159  &     5  &   $-$    &   $-$    &   $-$  &   0.115  &   0.041  &     7  &    97/02/22   &   98/04/03 \nl
 140  &      US   44           &  -0.045  &   0.025  &     9  &  -0.050  &   0.027  &     8  &  -0.094  &   0.031  &     5  &    97/02/11   &   97/12/27 \nl
 141  &      1E 0906+4254      &  -0.010  &   0.049  &    11  &  -0.267  &   0.070  &    12  &   0.058  &   0.081  &     7  &    96/12/30   &   98/03/02 \nl
 142  &      4C 05.38          &  -0.009  &   0.046  &     7  &  -0.133  &   0.140  &     6  &  -0.062  &   0.042  &     4  &    96/12/02   &   98/01/02 \nl
 143  &      MARK  704         &   0.036  &   0.025  &     3  &   0.048  &   0.021  &     3  &   0.111  &   0.036  &     4  &    96/02/03   &   97/12/24 \nl
 144  &      RXS J09189+3016   &  -0.185  &   0.132  &    10  &  -0.173  &   0.075  &     9  &  -0.228  &   0.078  &     9  &    97/02/20   &   98/01/26 \nl
 146  &      E 0917+341        &  -0.053  &   0.078  &     2  &  -0.032  &   0.073  &     2  &  -0.169  &   0.121  &     2  &    97/02/23   &   98/03/02 \nl
 148  &      PG 0923+201       &  -0.186  &   0.021  &     6  &  -0.257  &   0.030  &     4  &  -0.201  &   0.088  &     3  &    96/02/01   &   97/12/25 \nl
 149  &      MARK  705         &   0.093  &   0.019  &     8  &   0.503  &   0.038  &    13  &   0.221  &   0.017  &    10  &    96/02/06   &   97/12/24 \nl
 150  &      B2 0923+39        &   0.020  &   0.065  &     3  &  -0.293  &   0.106  &     1  &   $-$    &   $-$    &   $-$  &    96/12/21   &   98/03/15 \nl
 152  &      MS 09309+2128     &   $-$    &   $-$    &   $-$  &  -0.721  &   0.120  &     1  &  -0.473  &   0.160  &     1  &    96/11/30   &   97/12/27 \nl
 153  &      US  737           &  -0.016  &   0.034  &    13  &   0.167  &   0.046  &    13  &  -0.152  &   0.031  &    12  &    96/02/02   &   97/12/27 \nl
 154  &      MARK  707         &   $-$    &   $-$    &   $-$  &   0.199  &   0.047  &     2  &   0.120  &   0.048  &     3  &    96/02/03   &   97/12/24 \nl
\enddata
\end{deluxetable}

\begin{deluxetable}{clccccccccccc}
\scriptsize
\tablenum{1-Continued}
\tablecaption{}
\tablewidth{0pt}
\tablehead{
\colhead{Number} & \colhead{Name}   
& \colhead{$\Delta J$}   & \colhead{$\sigma_{\Delta J}$}  & \colhead{$n_J$}  
& \colhead{$\Delta H$}   & \colhead{$\sigma_{\Delta H}$}  & \colhead{$n_H$}   
& \colhead{$\Delta K'$}  & \colhead{$\sigma_{\Delta K'}$} & \colhead{$n_{K'}$}   
& \colhead{date1}   & \colhead{date2}
} 
\startdata
 155  &      TON 1078          &   0.033  &   0.041  &     5  &   0.025  &   0.057  &     4  &   0.051  &   0.087  &     1  &    96/02/02   &   98/01/05 \nl
 156  &      PG 0936+396       &  -0.122  &   0.064  &     7  &  -0.657  &   0.205  &     4  &  -0.398  &   0.239  &     3  &    96/02/02   &   98/02/12 \nl
 157  &      US  822           &   0.016  &   0.037  &     4  &   0.069  &   0.056  &     4  &   0.016  &   0.049  &     5  &    96/12/09   &   98/01/26 \nl
 159  &      HS 0940+4820      &   $-$    &   $-$    &   $-$  &   0.117  &   0.212  &     4  &   0.108  &   0.051  &     6  &    96/12/29   &   98/01/29 \nl
 160  &      2E 0944+4629      &   0.100  &   0.057  &     9  &  -0.043  &   0.084  &     8  &   0.072  &   0.077  &    11  &    97/01/02   &   98/02/10 \nl
 161  &      US  995           &   0.023  &   0.038  &     7  &   0.077  &   0.047  &     4  &   0.081  &   0.162  &     3  &    96/02/10   &   97/12/27 \nl
 162  &      HS 0946+4845      &  -0.322  &   0.049  &     3  &  -0.328  &   0.093  &     3  &   0.041  &   0.086  &     3  &    96/11/29   &   98/02/10 \nl
 164  &      US 1107           &  -0.470  &   0.044  &     2  &  -0.517  &   0.054  &     3  &  -0.673  &   0.123  &     2  &    96/11/29   &   98/01/16 \nl
 165  &      PG 0953+415       &  -0.128  &   0.024  &     7  &  -0.181  &   0.032  &     6  &  -0.226  &   0.030  &     4  &    96/02/02   &   97/12/18 \nl
 166  &      3C 232            &  -0.073  &   0.030  &     7  &  -0.054  &   0.049  &     4  &  -0.207  &   0.045  &     7  &    96/02/02   &   98/01/05 \nl
 167  &      NGC 3080          &   0.122  &   0.015  &    14  &   0.207  &   0.079  &    10  &   0.188  &   0.061  &     7  &    96/02/03   &   97/12/24 \nl
 168  &      IRAS 09595-0755   &  -0.555  &   0.018  &     8  &   $-$    &   $-$    &   $-$  &  -0.730  &   0.075  &     1  &    96/02/11   &   97/12/25 \nl
 170  &      KUV 10000+3255    &   0.248  &   0.272  &     4  &   $-$    &   $-$    &   $-$  &  -0.362  &   0.224  &     5  &    96/11/30   &   98/02/06 \nl
 172  &      PG 1001+05        &   0.015  &   0.074  &     8  &   0.288  &   0.066  &     5  &   0.188  &   0.029  &     5  &    96/02/10   &   97/12/24 \nl
 173  &      PKS 1004+13       &   0.393  &   0.153  &     5  &   0.026  &   0.052  &     4  &  -0.058  &   0.035  &     3  &    96/02/01   &   97/12/18 \nl
 174  &      RXS J10079+4918   &   0.205  &   0.051  &     3  &   0.147  &   0.161  &     3  &   0.277  &   0.130  &     3  &    97/02/20   &   98/01/19 \nl
 175  &      TON  488          &   0.733  &   0.072  &     3  &   $-$    &   $-$    &   $-$  &    $-$   &    $-$   &   $-$  &    96/01/30   &   96/04/05 \nl
 177  &      Q 1008+0058       &  -0.186  &   0.244  &     2  &   0.026  &   0.097  &     1  &  -0.112  &   0.137  &     2  &    96/02/10   &   96/11/30 \nl
 179  &      TON 1187          &  -0.050  &   0.030  &     8  &  -0.057  &   0.038  &     7  &   0.037  &   0.020  &     9  &    96/01/30   &   97/12/24 \nl
 180  &      PG 1011-040       &   0.053  &   0.021  &    12  &   0.139  &   0.025  &    11  &   0.235  &   0.029  &     6  &    96/02/11   &   97/12/25 \nl
 181  &      PKS 1011+23       &  -0.315  &   0.043  &     5  &  -0.283  &   0.052  &     6  &  -0.298  &   0.048  &     6  &    96/12/08   &   98/01/21 \nl
 182  &      PG 1012+008       &  -0.014  &   0.376  &     8  &  -0.114  &   0.038  &     5  &  -0.070  &   0.024  &     7  &    96/02/01   &   97/12/18 \nl
 190  &      Q 1047+067        &   $-$    &   $-$    &   $-$  &  -0.029  &   0.059  &     4  &   0.022  &   0.039  &     3  &    96/11/30   &   97/12/25 \nl
 191  &      MS 10470+3537     &   0.298  &   0.082  &     5  &  -0.222  &   0.090  &     4  &  -0.081  &   0.052  &     4  &    96/12/29   &   98/01/16 \nl
 193  &      PG 1049-005       &   0.045  &   0.043  &     8  &   0.002  &   0.050  &     9  &  -0.047  &   0.034  &     4  &    96/02/12   &   97/12/18 \nl
 194  &      MARK  634         &   0.074  &   0.073  &     1  &   $-$    &   $-$    &   $-$  &    $-$   &    $-$   &   $-$  &    96/02/03   &   97/12/25 \nl
 196  &      RXS J11008+2839   &   0.095  &   0.308  &     3  &   0.000  &   0.222  &     2  &   0.083  &   0.078  &     2  &    97/02/22   &   98/01/19 \nl
 197  &      MARK  728         &   0.174  &   0.039  &     9  &   0.062  &   0.053  &     4  &   0.270  &   0.039  &     5  &    96/02/03   &   97/12/25 \nl
 198  &      TOL 1059+105      &  -0.292  &   0.037  &     5  &  -0.319  &   0.262  &     5  &  -0.329  &   0.042  &     7  &    96/02/05   &   97/12/27 \nl
 199  &      1059.6+0157       &   $-$    &   $-$    &   $-$  &    $-$   &    $-$   &   $-$  &  -0.043  &   0.081  &     5  &    97/01/03   &   98/02/06 \nl
 200  &      PKS 1103-006      &   0.032  &   0.048  &     7  &   0.110  &   0.068  &     6  &  -0.074  &   0.066  &     5  &    96/02/12   &   98/01/05 \nl
 201  &      MC 1104+167       &  -0.815  &   0.036  &     8  &  -0.821  &   0.057  &     8  &  -0.666  &   0.045  &    12  &    96/02/02   &   98/01/19 \nl
 204  &      PG 1115+407       &  -0.154  &   0.021  &     7  &  -0.015  &   0.182  &     8  &  -0.130  &   0.074  &     8  &    96/02/12   &   98/04/03 \nl
\enddata
\end{deluxetable}

\begin{deluxetable}{clccccccccccc}
\scriptsize
\tablenum{1-Continued}
\tablecaption{}
\tablewidth{0pt}
\tablehead{
\colhead{Number} & \colhead{Name}   
& \colhead{$\Delta J$}   & \colhead{$\sigma_{\Delta J}$}  & \colhead{$n_J$}  
& \colhead{$\Delta H$}   & \colhead{$\sigma_{\Delta H}$}  & \colhead{$n_H$}   
& \colhead{$\Delta K'$}  & \colhead{$\sigma_{\Delta K'}$} & \colhead{$n_{K'}$}   
& \colhead{date1}   & \colhead{date2}
} 
\startdata
 205  &      PG 1116+215       &  -0.053  &   0.026  &     4  &   0.041  &   0.027  &     4  &   $-$    &   $-$    &   $-$  &    96/02/02   &   98/01/06 \nl
 206  &      MARK  734         &  -0.041  &   0.016  &     5  &  -0.013  &   0.027  &     6  &   0.038  &   0.084  &     3  &    96/02/05   &   97/12/24 \nl
 210  &      MARK  423         &   0.134  &   0.133  &     1  &   $-$    &   $-$    &   $-$  &    $-$   &    $-$   &   $-$  &    96/02/06   &   98/01/02 \nl
 211  &      US 2450           &  -0.007  &   0.036  &     8  &  -0.094  &   0.053  &     7  &   0.036  &   0.039  &     9  &    97/02/22   &   98/01/24 \nl
 212  &      MARK 1298         &  -0.174  &   0.211  &     5  &   $-$    &   $-$    &   $-$  &   0.213  &   0.203  &     3  &    96/02/06   &   98/01/24 \nl
 213  &      MARK 1447         &   0.114  &   0.038  &     6  &  -0.060  &   0.027  &     6  &  -0.083  &   0.028  &     6  &    96/02/05   &   97/12/24 \nl
 217  &      MCG  06.26.012    &   0.090  &   0.023  &     2  &   $-$    &   $-$    &   $-$  &   0.139  &   0.107  &     2  &    96/02/05   &   97/12/24 \nl
 218  &      MARK  744         &  -0.142  &   0.155  &     2  &   $-$    &   $-$    &   $-$  &  -0.091  &   0.186  &     1  &    96/02/06   &   98/01/24 \nl
 219  &      WAS  26           &  -0.086  &   0.030  &     9  &   0.552  &   0.076  &     7  &  -0.050  &   0.044  &     5  &    96/02/02   &   97/12/18 \nl
 220  &      CG  855           &  -0.234  &   0.039  &     3  &  -0.145  &   0.105  &     1  &  -0.170  &   0.133  &     1  &    96/02/05   &   98/01/02 \nl
 222  &      MC 1146+111       &   0.090  &   0.047  &     7  &   $-$    &   $-$    &   $-$  &  -0.209  &   0.139  &     5  &    96/12/08   &   98/03/07 \nl
 227  &      GQ COM            &  -0.244  &   0.048  &     7  &   0.103  &   0.043  &     8  &   0.083  &   0.027  &     6  &    96/01/31   &   97/12/18 \nl
 228  &      UGC  7064         &  -0.393  &   0.038  &     8  &  -0.065  &   0.022  &     3  &  -0.138  &   0.014  &     5  &    96/02/12   &   98/01/02 \nl
 230  &      PG 1211+143       &   0.158  &   0.043  &     4  &   0.153  &   0.029  &     3  &   0.052  &   0.012  &     3  &    96/01/31   &   98/01/24 \nl
 232  &      PKS 1216-010      &  -0.335  &   0.075  &     8  &  -0.401  &   0.104  &     4  &   $-$    &   $-$    &   $-$  &    96/02/02   &   98/01/21 \nl
 233  &      MARK 1320         &  -0.701  &   0.031  &     5  &  -0.551  &   0.036  &     5  &  -0.388  &   0.066  &     3  &    96/02/11   &   97/12/27 \nl
 235  &      Q 1220+0939       &   0.138  &   0.070  &     4  &  -0.183  &   0.143  &     4  &   $-$    &   $-$    &   $-$  &    96/12/21   &   98/01/30 \nl
 236  &      MS 12209+1601     &  -0.319  &   0.029  &     6  &  -0.386  &   0.044  &     8  &  -0.201  &   0.045  &     6  &    96/02/06   &   98/01/21 \nl
 238  &      2E 1224+0930      &  -0.129  &   0.052  &     2  &  -0.021  &   0.105  &     1  &  -0.114  &   0.226  &     2  &    97/01/20   &   98/01/31 \nl
 239  &      3C 273.0          &   0.320  &   0.024  &     5  &   0.378  &   0.013  &     4  &   0.251  &   0.256  &     7  &    96/01/30   &   98/01/24 \nl
 241  &      TON 1542          &  -0.031  &   0.071  &     5  &  -0.061  &   0.067  &     4  &   0.028  &   0.030  &     4  &    96/02/06   &   98/01/02 \nl
 242  &      CSO  150          &  -0.135  &   0.059  &     9  &   $-$    &   $-$    &   $-$  &  -0.009  &   0.082  &     4  &    96/01/30   &   96/02/21 \nl
 243  &      IC 3528           &   0.126  &   0.067  &     5  &   $-$    &   $-$    &   $-$  &  -1.787  &   0.222  &     1  &    96/02/06   &   97/12/25 \nl
 247  &      WAS 61            &  -0.598  &   0.018  &     5  &  -0.580  &   0.071  &     6  &  -0.490  &   0.020  &     8  &    96/02/11   &   98/01/02 \nl
 248  &      Q 1240+1546       &   0.077  &   0.113  &     3  &   0.180  &   0.102  &     1  &  -0.151  &   0.199  &     1  &    96/02/06   &   98/01/02 \nl
 249  &      CBS  63           &   $-$    &   $-$    &   $-$  &    $-$   &    $-$   &   $-$  &   0.021  &   0.137  &     2  &    97/02/23   &   98/02/12 \nl
 250  &      Q 1240+1746       &  -0.082  &   0.076  &     6  &  -0.163  &   0.106  &     7  &  -0.319  &   0.092  &     6  &    96/12/31   &   98/03/18 \nl
 251  &      4C 45.26          &   0.276  &   0.090  &     5  &   $-$    &   $-$    &   $-$  &   0.245  &   0.133  &     4  &    96/12/21   &   98/02/02 \nl
 257  &      3C 281            &   0.153  &   0.063  &     3  &  -0.033  &   0.085  &     3  &  -0.154  &   0.109  &     3  &    97/02/23   &   98/01/30 \nl
 260  &      CSO  835          &  -0.172  &   0.092  &     2  &   0.116  &   0.144  &     1  &   $-$    &   $-$    &   $-$  &    97/02/01   &   97/02/06 \nl
 262  &      PG 1309+355       &   0.020  &   0.128  &     4  &   1.578  &   0.488  &     4  &   0.124  &   0.114  &     2  &    96/02/02   &   98/03/01 \nl
 263  &      RXS J13129+2628   &   0.090  &   0.017  &     9  &   0.096  &   0.034  &     6  &   0.043  &   0.030  &     6  &    96/12/29   &   98/01/29 \nl
 264  &      Q 1316+0103       &  -0.131  &   0.082  &     7  &  -0.029  &   0.088  &     9  &   0.028  &   0.181  &     9  &    97/01/30   &   98/02/10 \nl
\enddata
\end{deluxetable}

\begin{deluxetable}{clccccccccccc}
\scriptsize
\tablenum{1-Continued}
\tablecaption{}
\tablewidth{0pt}
\tablehead{
\colhead{Number} & \colhead{Name}   
& \colhead{$\Delta J$}   & \colhead{$\sigma_{\Delta J}$}  & \colhead{$n_J$}  
& \colhead{$\Delta H$}   & \colhead{$\sigma_{\Delta H}$}  & \colhead{$n_H$}   
& \colhead{$\Delta K'$}  & \colhead{$\sigma_{\Delta K'}$} & \colhead{$n_{K'}$}   
& \colhead{date1}   & \colhead{date2}
} 
\startdata
 265  &      MARK 1347         &  -0.010  &   0.013  &    10  &  -0.059  &   0.014  &     9  &  -0.048  &   0.015  &     7  &    96/02/06   &   98/01/26 \nl
 266  &      Q 1326-0516       &  -0.070  &   0.026  &     5  &  -0.005  &   0.049  &     5  &   0.030  &   0.037  &     5  &    96/01/30   &   98/01/31 \nl
 267  &      MS 13285+3135     &   0.367  &   0.111  &     5  &   0.320  &   0.169  &     2  &   0.363  &   0.162  &     2  &    97/02/20   &   98/01/24 \nl
 271  &      IRAS 13349+2438   &  -0.079  &   0.013  &     7  &   0.035  &   0.057  &     3  &   0.106  &   0.079  &     2  &    96/02/10   &   98/01/24 \nl
 272  &      Q 1338-0030       &   $-$    &   $-$    &   $-$  &   0.144  &   0.101  &     1  &  -0.063  &   0.103  &     1  &    97/02/23   &   98/01/30 \nl
 273  &      TON  730          &   0.097  &   0.042  &     6  &   0.121  &   0.044  &     6  &   0.388  &   0.029  &     6  &    96/02/06   &   97/12/18 \nl
 274  &      MARK   69         &   0.023  &   0.027  &     5  &   0.063  &   0.266  &     4  &   $-$    &   $-$    &   $-$  &    96/02/05   &   98/03/02 \nl
 276  &      MARK  662         &  -0.084  &   0.099  &     2  &  -0.077  &   0.131  &     2  &   $-$    &   $-$    &   $-$  &    96/02/06   &   98/01/31 \nl
 278  &      MARK  463E        &   0.016  &   0.015  &     7  &   0.066  &   0.039  &     5  &  -0.056  &   0.129  &     7  &    96/02/06   &   98/01/31 \nl
 279  &      PG 1402+261       &   0.265  &   0.019  &     6  &   0.363  &   0.048  &     3  &   0.285  &   0.080  &     2  &    96/02/10   &   98/01/26 \nl
 280  &      PG 1404+226       &  -0.097  &   0.023  &     6  &  -0.147  &   0.034  &     5  &  -0.093  &   0.031  &     4  &    96/02/10   &   98/03/02 \nl
 281  &      OQ 208            &   0.126  &   0.016  &     8  &   0.302  &   0.058  &     6  &   0.128  &   0.156  &     1  &    96/02/06   &   98/01/26 \nl
 283  &      PG 1407+265       &  -0.055  &   0.038  &     5  &  -0.090  &   0.038  &     5  &   0.123  &   0.054  &     5  &    96/01/31   &   98/02/12 \nl
 284  &      PG 1411+442       &  -0.108  &   0.013  &     5  &  -0.071  &   0.014  &     6  &  -0.061  &   0.014  &     4  &    96/02/10   &   98/03/02 \nl
 285  &      PG 1415+451       &  -0.047  &   0.016  &    10  &  -0.059  &   0.022  &    10  &  -0.025  &   0.031  &     5  &    96/02/10   &   98/03/02 \nl
 286  &      NGC 5548          &   $-$    &   $-$    &   $-$  &   0.100  &   0.133  &     1  &   0.189  &   0.376  &     4  &    96/02/12   &   98/01/31 \nl
 287  &      H 1419+480        &  -0.175  &   0.015  &    11  &  -0.241  &   0.019  &    10  &  -0.297  &   0.014  &     9  &    96/02/12   &   98/03/02 \nl
 290  &      MARK  471         &   0.038  &   0.013  &     6  &   0.029  &   0.055  &     5  &   0.021  &   0.026  &     4  &    96/02/12   &   98/03/02 \nl
 291  &      B 1422+231        &   0.077  &   0.019  &     8  &   0.108  &   0.033  &     7  &   0.072  &   0.027  &     5  &    97/02/19   &   98/03/01 \nl
 292  &      2E 1423+2008      &   $-$    &   $-$    &   $-$  &    $-$   &    $-$   &   $-$  &   0.003  &   0.100  &     1  &    96/02/10   &   98/01/02 \nl
 293  &      MARK  813         &  -0.187  &   0.017  &     6  &  -0.216  &   0.026  &     6  &  -0.208  &   0.104  &     2  &    96/01/30   &   98/02/08 \nl
 294  &      B2 1425+26        &   0.021  &   0.060  &     6  &   0.203  &   0.338  &     3  &   $-$    &   $-$    &   $-$  &    96/01/30   &   98/02/12 \nl
 295  &      MARK 1383         &  -0.266  &   0.054  &     2  &  -0.081  &   0.013  &     3  &  -0.089  &   0.038  &     2  &    96/02/22   &   98/02/02 \nl
 296  &      MARK  684         &  -0.426  &   0.096  &    14  &  -0.010  &   0.016  &     9  &   0.028  &   0.024  &     5  &    96/02/06   &   98/01/26 \nl
 298  &      MARK  474         &   0.059  &   0.027  &    12  &   0.010  &   0.029  &     8  &   0.110  &   0.028  &     7  &    96/02/12   &   98/03/01 \nl
 300  &      MARK  478         &  -0.090  &   0.135  &    14  &  -0.126  &   0.013  &    10  &  -0.138  &   0.016  &    12  &    96/02/10   &   98/03/02 \nl
 301  &      PG 1444+407       &  -0.169  &   0.030  &     7  &  -0.026  &   0.027  &     8  &  -0.005  &   0.024  &     8  &    96/02/10   &   96/04/05 \nl
 302  &      Q 1446-0035       &   0.124  &   0.073  &     5  &   0.234  &   0.069  &     5  &   0.050  &   0.112  &     3  &    97/02/06   &   98/03/15 \nl
 303  &      PG 1448+273       &  -0.074  &   0.014  &    13  &  -0.061  &   0.017  &     8  &  -0.001  &   0.021  &     8  &    96/02/10   &   98/03/02 \nl
 304  &      MS 14564+2147     &  -0.407  &   0.025  &     9  &  -0.597  &   0.078  &    11  &  -0.657  &   0.041  &     6  &    96/02/12   &   98/03/02 \nl
 306  &      MARK  841         &   0.363  &   0.037  &     6  &   0.391  &   0.330  &     2  &   $-$    &   $-$    &   $-$  &    96/02/12   &   98/02/02 \nl
 308  &      PKS 1509+022      &   0.054  &   0.034  &    10  &   0.138  &   0.052  &    13  &   0.063  &   0.044  &     9  &    97/02/24   &   98/03/15 \nl
 309  &      MS 15198-0633     &  -0.265  &   0.020  &    14  &  -0.194  &   0.022  &    11  &  -0.270  &   0.018  &     9  &    96/02/22   &   98/03/07 \nl
\enddata
\end{deluxetable}

\begin{deluxetable}{clccccccccccc}
\scriptsize
\tablenum{1-Continued}
\tablecaption{}
\tablewidth{0pt}
\tablehead{
\colhead{Number} & \colhead{Name}   
& \colhead{$\Delta J$}   & \colhead{$\sigma_{\Delta J}$}  & \colhead{$n_J$}  
& \colhead{$\Delta H$}   & \colhead{$\sigma_{\Delta H}$}  & \colhead{$n_H$}   
& \colhead{$\Delta K'$}  & \colhead{$\sigma_{\Delta K'}$} & \colhead{$n_{K'}$}   
& \colhead{date1}   & \colhead{date2}
} 
\startdata
 310  &      LB 9695                   &  -0.284  &   0.085  &    11  &   $-$    &   $-$    &   $-$  &    $-$   &    $-$   &   $-$  &    97/01/30   &   98/03/17 \nl
 311  &      OR 139                    &  -0.086  &   0.061  &     7  &   0.039  &   0.063  &     7  &   0.100  &   0.091  &     8  &    97/01/28   &   98/02/06 \nl
 312  &      QNZ5:02                   &   0.003  &   0.054  &     9  &   0.023  &   0.070  &     8  &  -0.037  &   0.046  &     6  &    97/02/23   &   98/02/12 \nl
 313  &      MARK 1098                 &  -0.100  &   0.012  &    13  &  -0.136  &   0.015  &    12  &  -0.135  &   0.018  &     5  &    96/02/10   &   98/02/02 \nl
 314  &      NGC 5940                  &  -0.153  &   0.014  &    14  &  -0.237  &   0.139  &     9  &  -0.123  &   0.019  &     4  &    96/02/22   &   98/01/31 \nl
 315  &      KUV 15524+2153            &  -0.082  &   0.038  &    11  &  -0.014  &   0.038  &    13  &  -0.081  &   0.043  &    10  &    97/02/22   &   98/03/16 \nl
 318  &      PG 1634+706               &   0.074  &   0.011  &     9  &   0.055  &   0.017  &     9  &   $-$    &   $-$    &   $-$  &    97/02/22   &   98/03/07 \nl
 319  &      RXS J16446+2619           &  -0.042  &   0.080  &     8  &  -0.117  &   0.055  &     9  &   0.127  &   0.308  &     8  &    97/02/11   &   98/03/16 \nl
 320  &      TEX 1652+151              &  -0.130  &   0.047  &    15  &   0.100  &   0.093  &    14  &   0.041  &   0.042  &    13  &    97/02/20   &   98/03/16 \nl
 321  &      2E 1654+3514              &   $-$    &   $-$    &   $-$  &   0.215  &   0.133  &     8  &  -0.067  &   0.108  &     8  &    97/02/20   &   98/03/18 \nl
 323  &      PKS 1739+18C              &  -0.141  &   0.029  &    24  &   0.167  &   0.031  &    22  &   0.095  &   0.024  &    13  &    97/02/24   &   98/03/07 \nl
 328  &      3C 459.0                  &  -0.085  &   0.063  &     4  &   0.248  &   0.102  &     5  &  -0.014  &   0.062  &     3  &    96/11/25   &   97/12/31 \nl
 330  &      PB 5577                   &  -0.192  &   0.098  &     5  &   0.047  &   0.116  &     3  &  -0.079  &   0.113  &     3  &    96/12/02   &   97/12/27 \nl
 331  &      Q 2352+0025               &   0.052  &   0.048  &     8  &   0.033  &   0.053  &     7  &   0.022  &   0.035  &     6  &    96/12/23   &   97/12/31 \nl
\enddata
\tablecomments{The sequential number in the first column is the same
   as in the object list in Paper I. The numbers $n_J$, $n_H$, 
   and $n_{K'}$ represent those of reference objects used in the 
   differential photometry in the $J$, $H$, and $K'$ bands,
   respectively.
}
\end{deluxetable}

\clearpage


\clearpage

\begin{deluxetable}{llccclccclccc}
\scriptsize
\tablecaption{Ratio of varied AGNs with $2\sigma$ and $3\sigma$ confidence.}
\tablewidth{0pt}
\tablehead{
   \hspace{15mm} Sample   & &&\colhead{$J$}  & & & &\colhead{$H$}  & &  & &\colhead{$K'$} & 
\nl \cline{3-5} \cline{7-9} \cline{11-13}
   &  & \colhead{$>2\sigma$} & \colhead{$>3\sigma$}  & $n$ &
   & \colhead{$>2\sigma$} & \colhead{$>3\sigma$}  & $n$ & 
   & \colhead{$>2\sigma$} & \colhead{$>3\sigma$}  & $n$ 
}
  
\startdata
all AGNs . . . . . . . . . . . . . . . . .& & 61\%   & 47\%   & 165 & &  58\%   & 42\%   & 134 & & 56\%   & 42\%   & 135 \nl
\hspace{8mm} $\sigma_{\Delta{m_{AGN}}} < 0.03$ . . . . . . . .& & 85\% & 73\% & 52 & & 82\% & 67\% & 33 & & 73\% & 65\% & 37   \nl
\hspace{8mm} $0.03 < \sigma_{\Delta{m_{AGN}}} < 0.05 $ . . . .& & 59\% & 44\% & 54 & & 68\% & 50\% & 34 & & 60\% & 43\% & 47   \nl
\hspace{8mm} $0.05 < \sigma_{\Delta{m_{AGN}}} < 0.10 $ . . . .& & 41\% & 27\% & 59 & & 40\% & 27\% & 67 & & 41\% & 26\% & 51   \nl
   \nl
radio quiet  AGNs . . . . . . . . . . . . & & 59\%   & 47\%   & 116 & &  60\%   & 42\%   & 97 &  & 59\%   & 41\%   & 100 \nl
\hspace{8mm} $\Delta t_{rest}=100-400 days$ . . . .& & 38\%   & 27\%   & 49  & &  32\%   & 11\%   & 38 & &  41\%   & 14\%   & 42  \nl
\hspace{8mm} $\Delta t_{rest}=400-800 days$ . . . .& & 75\%   & 61\%   & 67  & &  78\%   & 63\%   & 59 & &  72\%   & 60\%   & 58  \nl
   \nl
radio loud  AGNs . . . . . . . . . . . . & & 67\%   & 48\%   & 42  & &  50\%   & 41\%   & 32 & &  44\%   & 41\%   & 32 \nl
\hspace{8mm} $\Delta t_{rest}=100-400 days$ . . . .& & 60\%   & 43\%   & 30  & &  48\%   & 38\%   & 21 & &  36\%   & 32\%   & 25 \nl
\hspace{8mm} $\Delta t_{rest}=400-800 days$ . . . .& & 83\%   & 58\%   & 12  & &  55\%   & 46\%   & 11 & &  71\%   & 71\%   & 7  \nl
\enddata
\tablecomments{
  The number $n$ represents that of AGNs in each sample.
}
\end{deluxetable}

\end{document}